\documentclass[a4paper,fleqn,usenatbib]{mnras}

\usepackage{newtxtext,newtxmath}
\usepackage{natbib}
\usepackage{mathptmx}
\usepackage[T1]{fontenc}
\usepackage{ae,aecompl}

\usepackage[pdftex]{graphicx}
\usepackage{xfrac}
\usepackage{geometry}
\usepackage{amsmath}	% Advanced maths commands
\usepackage{amssymb}	% Extra maths symbols
\usepackage{pdflscape}
\usepackage{multicol}
\usepackage{multirow}
\usepackage{caption}
\usepackage{titlesec}
\usepackage[flushleft]{threeparttable}
\title[A high-resolution view of the jets in 3C\,465]{A high-resolution view of the jets in 3C\,465}

\author[E. Bempong-Manful et al.]{E. Bempong-Manful$^{1,2}$\thanks{E-mail: e.bempong-manful@bristol.ac.uk (EBM)}, M.J. Hardcastle$^{1}$, M. Birkinshaw$^{2}$, R.A. Laing$^{3}$, J.P. Leahy$^{4}$, \newauthor D.M. Worrall$^{2}$
\\
$^{1}$School of Physics, Astronomy and Mathematics, University of Hertfordshire, College Lane, Hatfield AL10 9AB, UK 
\\
$^{2}$School of Physics, University of Bristol, Tyndall Avenue, Bristol BS8 1TL, UK
\\
$^{3}$SKA Organisation, Jodrell Bank Observatory, Lower Withington, Macclesfield, Cheshire SK11 9DL, UK
\\
$^{4}$Jodrell Bank Centre for Astrophysics, Alan Turing Building, School of Physics and Astronomy, University of Manchester,\\ Manchester, M13 9PL, UK\\}

\date{Accepted 2020 May 17. Received 2020 May 15; in original form 2020 March 31}

\pubyear{2020}

\begin{document}
\label{firstpage}
\pagerange{\pageref{firstpage}--\pageref{lastpage}}
\maketitle

% Abstract of the paper
\begin{abstract}
We present new high-resolution and high-sensitivity studies of the jets in the WAT source 3C\,465, using deep transverse-resolved radio observations from e-MERLIN, and with complementary observations from the VLA. We derive a lower limit $\beta_{\rm j}$ = ($\nu_{\rm j}$/$c$) $\gtrsim$ 0.5 for the jet speed, and an upper limit $\theta_{\rm j}$ $\lesssim$ 61$^{\circ}$ for the jet angle to the line of sight. The jet spectral index ($\alpha$, defined in the sense $S \propto \nu^{\alpha}$) is fairly constant (<$\alpha_{\rm jet}$> = $-$0.7), and spectral flattening within 4.4 kpc of the core coincides with bright knots and is consistent with the site of X-ray particle acceleration at the base of the radio jet found in previous studies. There is little difference between the spectra of the two hotspot components, plausibly indicating that electron populations of the same properties are injected there. The NW and SE plumes are approximately homologous structures, with variations in mass injection and propagation in external pressure and density gradients in the two regions plausibly accounting for the slightly steeper spectrum in the NW plume, <$\alpha_{\rm NWp}$> = $-$1.43 compared with the SE plume, <$\alpha_{\rm SEp}$> = $-$1.38. Our synchrotron lifetime model supports plausible reacceleration of particles within the plume materials. Overall, our results show that the first-order Fermi process at mildly relativistic and non-relativistic shocks is the most likely acceleration mechanism at play in 3C\,465 and distinguish differences between the acceleration at $\beta_{\rm j}$ $>$ 0.5 and $\beta_{\rm j}$ $<$ 0.5. The former case can accelerate electrons to higher Lorentz factors. 
\end{abstract}

% Select between one and six entries from the list of approved keywords.
% Don't make up new ones.

%%%%%%%%%%%%%%%%%%%%%%%%%%%%%%%%%%%%%%%%%%%%%%%%%%

%%%%%%%%%%%%%%%%% BODY OF PAPER %%%%%%%%%%%%%%%%%%

\section{Introduction}

Relativistic plasma ejected from SMBHs at the centres of massive galaxies is known to play a key role in the AGN feedback cycle, and consequently the formation and evolution of structure in the Universe. The formation, collimation and acceleration of AGN outflows, notwithstanding the vast range of physical systems from which they are produced, are thought to involve essentially similar physical mechanisms (e.g. \citealt{Wiita01}). In the case of a supersonic beam, dynamical instabilities associated with these outflows are rich enough to allow the formation of structures such as knots, filaments and wiggles (e.g. \citealt{Stone97}) and these have been observed and studied in detail (e.g. \citealt{Hardcastle02+,Laing06}). 

\begin{table*}
	\begin{threeparttable}
		\caption{Summary of observations}
		\label{tab:table1}
			\begin{tabular}{|l|c|c|c|c|c|c|c|c|c|r|} 
				\hline
				Data  &\multicolumn{2}{c}{e-MERLIN Observations}  &&\multicolumn{4}{c}{VLA Observations}\\
				& Epoch 1 &Epoch 2 && A& B& C& D\\
				(1)& (2)&(3) && (4)&(5)&(6)&(7)\\
				\hline
				Project ID & EGJ\_20150412 &EGJ\_20150413 && 12A-195& 12A-195\\ 
				& & && --$^\textbf{*}$ &-- &-- &-- \\
				Central Freqreuncy (GHz) & 1.51& 1.51 && 1.50& 1.50\\ 
				& & && 8.44$^\textbf{*}$ &8.49 &8.00  &8.45 \\
				Total Bandwidth (MHz) & 512 &512 && 512 &512 \\ 
				& & && 25$^\textbf{*}$ &50 &50 &50 \\
				Time on Source (hours) & 19.0 &19.3 && 1.2 &1.2 \\
				& & && 2.0$^\textbf{*}$ &1.5 &6.7 &1.3 \\
				Date & 12 Apr 2015 &13 Apr 2015 && 31 Oct 2012 &28 May 2012 \\ 
				& & && 15 Jan 2001$^\textbf{*}$ &19 May 2001 &20 Sep 1989 &01 Dec 1989 \\
				\hline
			\end{tabular}

		\begin{tablenotes}
		\small \item{Notes: $^\textbf{*}$denote X-band ($\sim$ 8.5 GHz) VLA observations of 3C\,465. Columns (1), (2) \& (3) are self explanatory. Columns (4), (5), (6) \& (7) represent the VLA observing configurations used in the present study. The VLA X-band observations were made by \citet{Hardcastle&Sakelliou04}$^{A,B}$ and \citet{Eilek&Owen02}$^{C,D}$.}	\end{tablenotes}
	\end{threeparttable}
\end{table*}

\par In spite of the complex morphologies exhibited by extragalactic radio sources, both types I and II of the Fanaroff and Riley (\citealt{Fanaroff&Riley74}) morphological classification of radio-loud AGNs tend to have pairs of jets near the plane of the sky (e.g. \citealt{Urry&Padovani}). The FR Is are relatively low-power radio sources with twin jets that are relativistic on parsec scales, but decelerate to sub-relativistic speeds on kpc scales. The jets exhibit considerable brightness asymmetry at their base (e.g. \citealt{Laing99+}), are thought to be transonic (i.e., no strong shocks are observed at their termination points) on kpc scales, and both jets are easily observed on large scales due to the absence of strong Doppler boosting effects (e.g. \citealt{Worrall07+}). \citet{Bridle&Perley84} also notes that the apparent magnetic field in these FR I jets changes from longitudinal to transverse as the jet propagates.
\par Wide-angle tailed (WAT) radio galaxies, generally classified as FR I sources, are associated with central cluster galaxies (e.g. \citealt{Owen&Rudnick76}) and have luminosities at the FR I/FR II break. They exhibit one or two well-collimated jets which usually extend for tens of kpc before flaring into characteristic plumes at their termination point, and the jets have polarisation structures that closely resemble those of FR II sources (e.g. \citealt{O'Donoghue93+, Hardcastle&Sakelliou04}). Although they form a small minority of the radio source population, the distinctive bending of the tails that gave the class its name has long been of interest (e.g. \citealt{Eilek84+}). Numerous theories including electrodynamic effects (\citealt{Bodo85+}), gravitational bending (\citealt{Burns82+}), buoyancy effects (\citealt{Worrall95+}) and ram pressure (\citealt{Venkatesan94+}) have been proposed in the literature to account for tail-bending in radio galaxies. However, WATs with their varied bending angles (i.e., 30 -- 115 degrees) (e.g. \citealt{O'Donoghue93+}) and collimated jets which often travel several kpc into the ICM before flaring into plumes, present an unusual challenge. Earlier work by \citet{Hardcastle&Sakelliou04} in a study of selected WATs in Abell clusters suggests that the jets in WATs terminate in a variety of ways, further underscoring the complexity of the physical mechanism that initiates the bending in this class of radio sources. The tails of WAT sources generally bend in a common direction, resulting in their overall characteristic U, V, or C shape. These sources are generally assumed to be shaped by motion of the host galaxy relative to the cluster: the host galaxies are thought to be nearly stationary at the bottom of cluster potential wells, moving with velocities $\sim$ 200 km s$^{-1}$ in an oscillatory motion of amplitude $<$ 0.3 of a core radius (e.g. \citealt{Burns82+}).

\par 3C\,465 is associated with NGC7720, the dominant {\textquotedblleft{diffuse}\textquotedblright} galaxy in the cluster Abell 2634, and is among the best studied WAT sources in the northern sky due to its proximity and peculiar morphology within this class of radio sources. Detailed imaging studies in the optical (e.g. \citealt{Colina&Perez90, Capetti05+}), radio (e.g. \citealt{Leahy84, Hardcastle&Sakelliou04}) and X-ray (e.g. \citealt{Schindler&Prieto97, Hardcastle05+}) have previously been made. Its jet and hotspots as shown in our total intensity maps (see Section \ref{sec:3.3}) have been extensively studied and large-scale properties of the plumes are also well known. 
\par Although driven by essentially similar underlying physical mechanisms, the morphology, kinematics and dynamics of AGN jets are heavily influenced by differences in host galaxy properties and environment, and efforts are ongoing to constrain the physics driving the jet structures in these cosmic outflows, including where and how particles are accelerated in the jets and hotspots. As part of the e-MERLIN Legacy project, which aims to resolve some of these key questions in extragalactic jet physics, we present new high-resolution and high-sensitivity studies from multi-configuration, multi-frequency VLA and e-MERLIN observations of the WAT source 3C\,465.  
\par Throughout this paper, we assume a concordance cosmology with $\Omega_{\rm m}$ = 0.27, $\Omega_\Lambda$ = 0.73, and $H_0$ = 75 km s$^{-1}$ Mpc$^{-1}$. At the current best known redshift of 3C\,465 -- i.e., $z$ = 0.03035$\pm$0.00015 (\citealt{Smith04+}), one arcsec is equivalent to a projected length of 0.56 kpc. Spectral indices, $\alpha$ are defined in the sense $S \propto \nu^{\alpha}$. J2000 co-ordinates are used throughout.
 
\begin{table*}
	\begin{threeparttable}
		\caption{Properties of radio maps presented}
		\label{tab:table2}
		\begin{tabular}{|l|c|c|c|c|c|c|c|c|r|} 
			\hline
			Map  
			&\multicolumn{3}{c}{Restoring beam} 
			&\multicolumn{1}{c}{Core flux} 
			&\multicolumn{1}{c}{Off-source}
			&\multicolumn{1}{c}{Figure}\\
			&Major axis &Minor axis &Pos. angle &density (mJy) &noise ($\mu$Jy/beam) &label\\
			&(arcsec) &(arcsec) &($^\circ$)\\
			\hline
			VLA$^{L}$ &1.37 &1.13 &71.0 &214 {$\pm$} 4 &34 &1\\
			
			e-MERLIN$^{L}$ &0.26 &0.13 &19.3 &207 {$\pm$} 1 &31 &2\\
			
			e-MERLIN$^{L}$+VLA$^{L}$ &0.31 &0.16 &18.9 &207 {$\pm$} 2 &33 &3\\
			
			VLA$^{L}$ &1.5 &1.5 &0.00 &217 {$\pm$} 3 &34 &NS\\
			
			VLA$^{X}$ &1.5 &1.5 &0.00 &214 {$\pm$} 1 &20 &NS\\
			
			VLA$^{X}$ &0.5 &0.5 &0.00 &213 {$\pm$} 1 &22 &NS\\
			
			e-MERLIN$^{L}$+VLA$^{L}$ &0.5 &0.5 &0.00 &218 {$\pm$} 7 &31 &NS\\
			\hline
		\end{tabular}
		\begin{tablenotes}
			\small \item{Notes: The superscripts L and X respectively denote $\sim$ 1.5 GHz and $\sim$ 8.5 GHz observing frequency used in our present study. Maps not shown in the text are denoted NS in the figure label column.}
		\end{tablenotes}
	\end{threeparttable}
\end{table*}
 
\section{Data and Methods}

\subsection{Observations, calibration and imaging}
\label{sec:2.1}   

3C\,465 was observed with both the Karl G. Jansky Very Large Array (VLA) and e-MERLIN at L-band continuum frequencies centred on 1.5 GHz using a bandwidth of 512 MHz. The bright sources 3C\,48 and 3C\,286 were used as the flux density calibrators for the VLA and e-MERLIN observations respectively. Our VLA data had 16 adjacent spectral windows -- each 64 MHz spectral window had 64 channels 1.0 MHz wide with 1 second and 3 seconds of integration time for A and B-configurations respectively; and our e-MERLIN data had 8 adjacent intermediate-frequencies (IF) -- each IF had 512 channels which was later averaged to 128 channels per IF with 2 seconds integration time. L-band observations in two VLA configurations (see Table \ref{tab:table1}) were undertaken in view of the scientific objectives of the present study -- i.e., the e-MERLIN array with its long baselines yielded higher angular resolution of the science target, whereas the more diffuse large scale structure was sampled by the shorter baselines of the VLA to deliver the best possible $uv$ coverage, which is essential to constructing high fidelity images. The data were calibrated, imaged and self-calibrated using standard procedures in CASA and AIPS respectively for the VLA and e-MERLIN datasets. As usual for VLBI observations, we performed fringe-fitting on the e-MERLIN datasets using the AIPS task [FRING] prior to calibration. 
\par We do not expect large errors from the transfer of the flux-density scale over our frequency range since this was done with the primary and secondary calibrators observed at similar elevations. However, for accuracy and reliability, we used values of flux density and visibility models for 3C\,286 provided with the AIPS package for our e-MERLIN datasets; and, in the case of the VLA, we rescaled the assumed flux in the initial data reduction steps by multiplying the ratio of the flux density of the primary calibrator given by \citet{Perley&Butler13}. The estimated uncertainty in the absolute flux density scale at 1.5 GHz is $\approx$ 2$\%$. Image deconvolution was executed using the conventional (single-resolution) CLEAN algorithm as implemented in CASA and AIPS. After obtaining a sufficiently good image of our science target from the first round of deconvolution, we embarked on the so called \textit{self-calibration} process -- i.e., we used the obtained model image of the target to solve for new and improved complex gain values of the visibilities, and then re-applied the new solutions to the science target, and repeated the deconvolution process. We performed several rounds of self-calibration, each time varying the \textit{solution interval} over which the complex gains were derived until we observed no significant decrease in the image noise after consecutive cycles. We compared our images with the X-Band ($\sim$ 8.5 GHz) VLA data of \citet{Hardcastle&Sakelliou04} from all four configurations (A, B, C and D) of the VLA with bandwidths of 25 MHz (for the A-configuration observations) and 50 MHz (for the rest).

\begin{figure*}\setlength\intextsep{-1pt}
	\centering
	\includegraphics[width=0.9\textwidth,height=0.9\textheight,keepaspectratio=true]{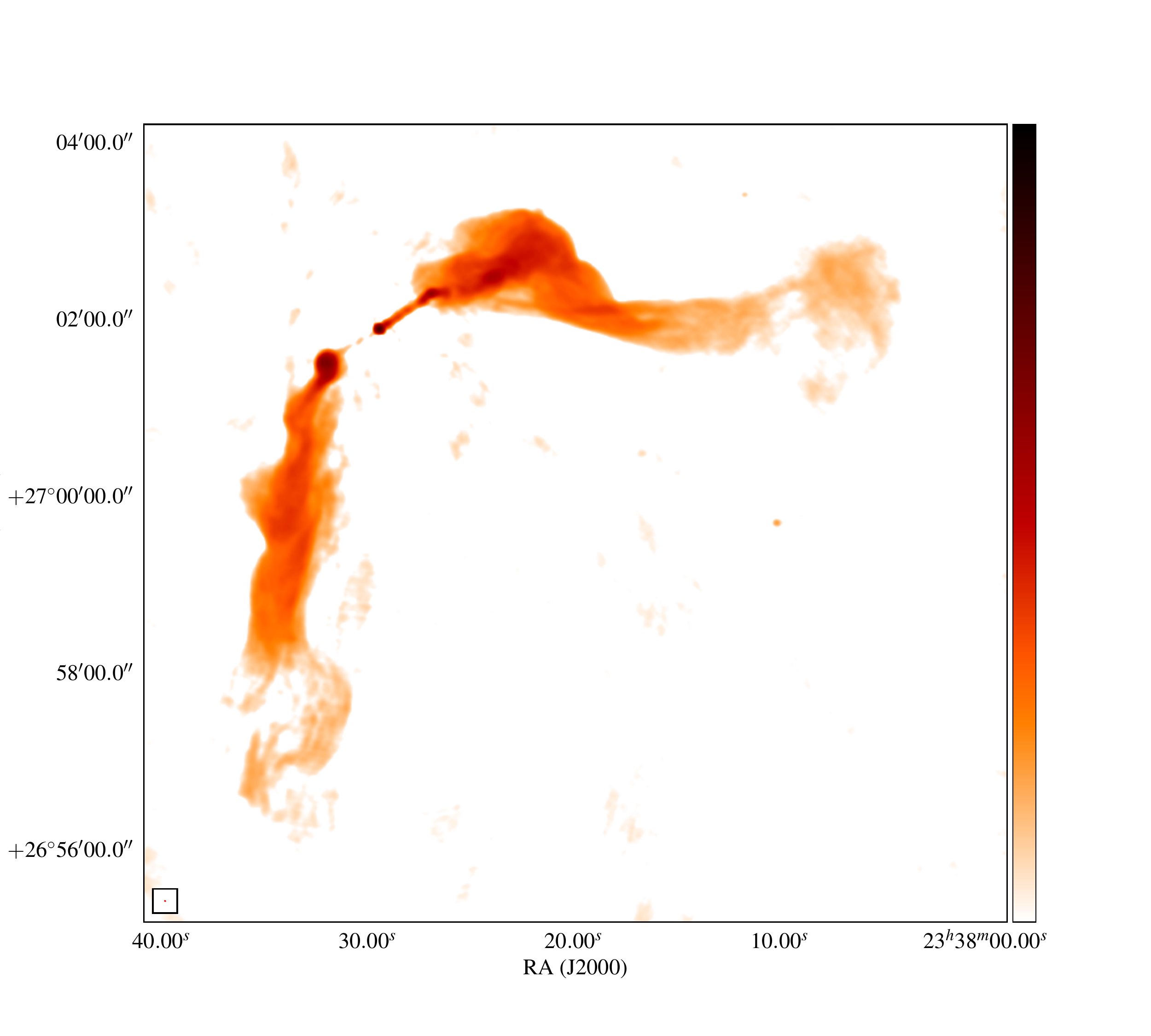}
	\vspace*{-0.75cm}
	\caption{$1.37\times1.13$-arcsec resolution radio map of the WAT radio galaxy 3C\,465 observed in A and B--configurations of the expanded VLA at L-Band ($\sim$1.5 GHz) continuum frequencies -- showing the peaked central emission and the elongation of the overall radio structure in the NW-SE direction.}
	\label{fig:vlamap}
\end{figure*}

\begin{figure*}
	\centering
	\includegraphics[width=0.9\textwidth,height=0.9\textheight,keepaspectratio=true]{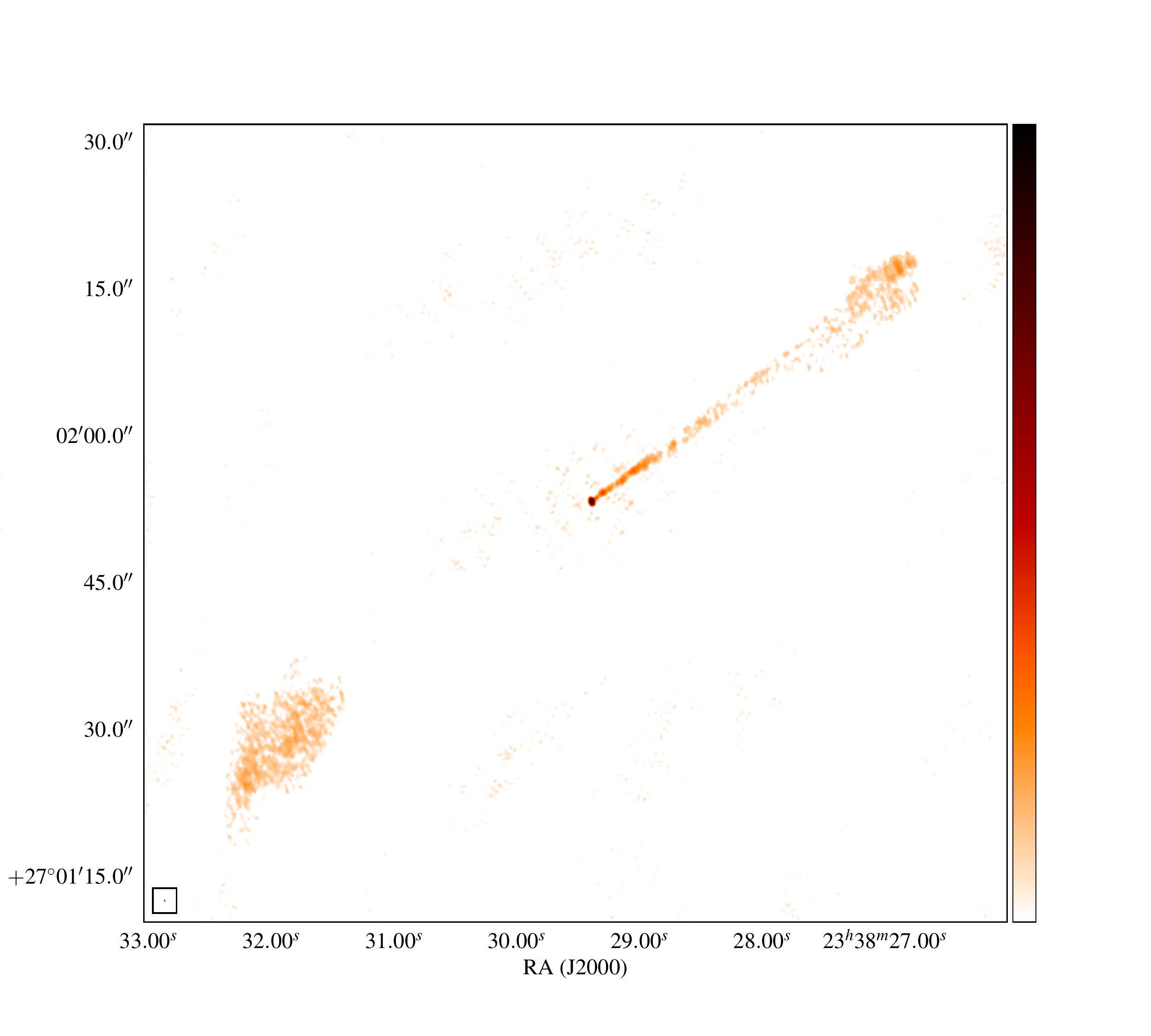}
	\vspace*{-0.75cm}
	\caption{$0.27\times0.15$-arcsec resolution radio map of the WAT source 3C\,465 observed at L-Band ($\sim$1.5 GHz) continuum frequencies with the e-MERLIN array -- showing the sidedness of the radio jet (note the visible bright knotty structures at the jet base and the elliptical shape of the recovered primary beam)}
	\label{fig:emlnmap}
\end{figure*} 

\subsection{Data combination and mapping}
\label{sec:2.2} 

Data combination from different interferometers -- as in the present study -- can be problematic due to fundamental differences in the distributions of data points in the $uv$ plane and errors in cross-calibration of the visibilities from the different arrays. We have carried out the combination in the $uv$ plane to allow for better constraints on the CLEAN algorithm (see \citealt{Biggs&Ivison08}). 
\par We calibrated and imaged [using the \textit{Multi-scale} and \textit{Multi-frequency synthesis} parameters in the CLEAN task] separately the A and B-configurations of the VLA datasets before combining the two configurations into a single measurement set. A similar approach was used for the two e-MERLIN observing epochs which included Lovell and non-Lovell baselines respectively. The resulting datasets were then used to create a complete e-MERLIN/VLA dataset, with appropriate weighting for the two array visibilities to account for differences in the weighting schemes for the two arrays. Finding the optimum weighting factor requires verification by visual inspection, and in the present study we varied the weights and formed the dirty beam using \textit{wsclean} (see \citealt{Offringa14+}) to find the point at which the baselines of the combined dataset were not dominated by the VLA. Such comparable statistical weightings were obtained with weighting factor

\begin{equation}
\frac{W_{\rm VLA}}{W_{\rm eMERLIN}} = \frac{3\times10^{-7}}{1}
\label{eqn1}
\end{equation}
This allowed us to construct total intensity maps of the combined dataset with a beam size close to the e-MERLIN -- only beam ($\sim$ 150 mas). Deconvolution of the e-MERLIN/VLA dataset was executed in \textit{wsclean} with the \textit{Briggs} robustness parameter set to $-$1.5 to give further weight at the long baselines contributed by e-MERLIN. 

\subsection{Resolution matching}
\label{sec:2.3} 

Our combined e-MERLIN/VLA map (hereafter referred to as the 1.5 GHz map) has a resolution of $\sim$ $0.31$ arcsec $\times$ $0.16$ arcsec. It would have been possible to obtain a beam size approximately equal to the e-MERLIN only beam size. However, it is impossible to obtain good image fidelity of the 8.5 GHz map at such high spatial resolutions. For accurate determination of spectral indices across the two frequencies, we needed to convolve the 1.5 and 8.5 GHz maps to equivalent resolutions. This allowed us to effectively measure flux densities from the same spatial region from each map. At higher resolutions we are less sensitive to extended emission and so we use an 0.5-arcsec resolution map for the jet and hotspot regions and 1.5-arcsec resolution map for the plumes. We have performed primary beam correction on the 8.5 GHz map. This allows us to measure precisely matched flux densities and corresponding errors since the shortest baselines sampled by the two maps are similar and we are sensitive to the same extended structure (plumes) across the two frequencies. 
\vspace{-10pt}
\section{The WAT source 3C\,465}

\subsection{The radio core}
\label{sec:3.2} 

There is no observed variability of the radio core of 3C\,465 over the time-scales of the VLA observations at 1.5 or 8.5 GHz and/or e-MERLIN observations at 1.5 GHz, within the errors imposed by the uncertainty of absolute flux calibration at either the VLA or e-MERLIN array. Its flux density measured from the two interferometry (and in the case of the VLA, the two configurations deployed in our observations) is reported in Table \ref{tab:table2}. The best position for the core is RA 23h38m29.393s, Dec 27$^{\circ}$01$'$53.$''$25.   

\begin{figure*}
	\centering
	\includegraphics[width=0.9\textwidth,height=0.9\textheight,keepaspectratio=true]{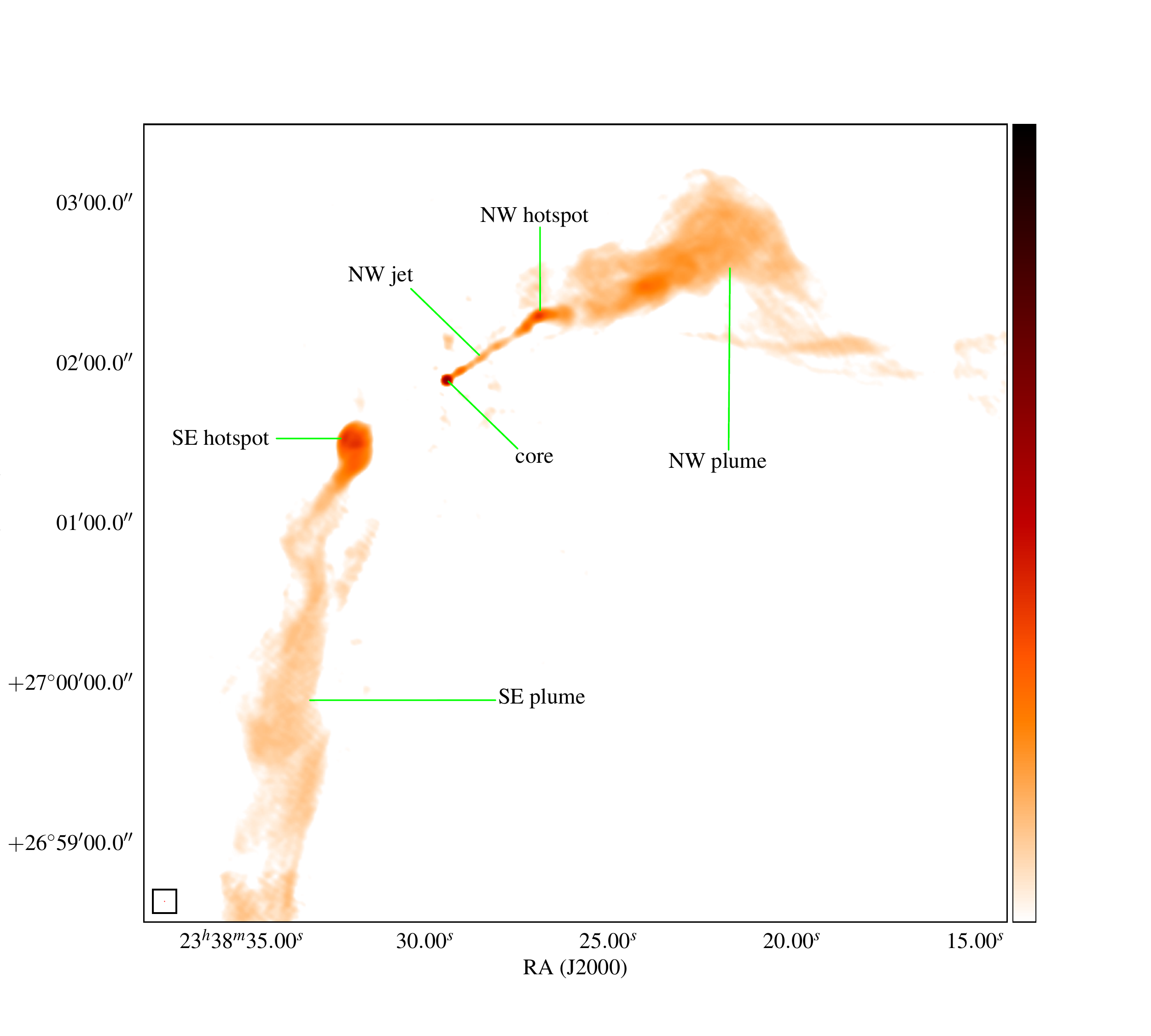}
	\vspace*{-0.75cm}
	\caption{Combined L-Band ($\sim$1.5 GHz) e-MERLIN plus VLA map at $0.31\times0.16$-arcsec resolution (note the continued collimated outflow from the NW hotspot into the base of the NW plume).}
	\label{fig:emlnvlamap}
\end{figure*} 
\subsection{Jets, knots, hotspots and plumes}
\label{sec:3.3} 

Our low-resolution map (Fig. \ref{fig:vlamap}) shows the overall radio structure, which is consistent with the characteristic U or C-shaped morphology exhibited by the WAT class of radio sources. The total angular extent of 3C\,465 estimated from our 1.5 GHz VLA map is 7.6 arcmin. This corresponds to a projected linear size of $\sim$ 275 kpc -- about 30 $\%$ smaller than that measured by \citet{Leahy96+} using shorter VLA baselines. Our radio image shows a narrow, well collimated jet (denoted NW; Fig. \ref{fig:emlnvlamap}) emanating from the core of the galaxy; this jet which is presumably pointing towards us exhibit considerably high surface brightness over almost its full length. In our low-resolution map (Fig. \ref{fig:vlamap}) a faint counterjet can be seen. Our e-MERLIN only image (Fig. \ref{fig:emlnmap}) represent the highest resolution \textit{deep transverse-resolved radio map} of 3C\,465 to date, and with the short baseline contribution from the VLA in our combined e-MERLIN/VLA map (Fig. \ref{fig:emlnvlamap}) we can further construct the highest resolution and sensitivity map of the source. In our high resolution map (Fig. \ref{fig:emlnmap}) the counterjet is not visible. 
\par The NW jet is well resolved and has a deconvolved cross-sectional width of $\sim$ 0.32 arcsec measured over the inner region where the jet is bright and straight. At high resolution, there is clear evidence of a knotty structure at the base of the NW jet of 3C\,465 -- a feature observed earlier by \citet{Hardcastle&Sakelliou04}. However at e-MERLIN resolution (see Fig. \ref{fig:emlnmap}) our good image fidelity allows us to infer multiple bright knotty structures -- two of which are visible at the edge of the core at $\sim$ 38$^{\circ}$(209 pc) and $\sim$ 38$^{\circ}$(330 pc), the point where the jet is first seen emanating from the central engine, and a third knot which lies between the two knots denoted NJ1 by \citet{Hardcastle&Sakelliou04}. Beyond the inner jet there is no clear evidence of such knotty structures until the main site of energy dissipation in the bright compact region (denoted NW hotspot) near the base of the NW plume (Fig. \ref{fig:emlnvlamap}).
\par This NW hotspot is at $\sim$ 23.4 kpc projected from the core and equivalent in projection from the core to the SE hotspot. The SE hotspot exhibits a relatively broad bright structure compared with the NW hotspot. Since both jets must carry power of equal magnitude (for momentum conservation), it is difficult to assign a physical interpretation to this striking asymmetry between the two hotspot regions which appears characteristic of this class of radio galaxies; see \citet{Hardcastle&Sakelliou04} for similar asymmetry in a sample of 7 WAT sources including 3C\,465. A plausible explanation however could be the degree of jet-environment interactions at the two sites -- with what appears to be a ``mini'' overpressured cocoon at the SE hotspot region resulting from a relatively higher density environment in pressure equilibrium with the ICM. 
\par From the NW hotspot, the jet is observed to propagate further without disruption up to $\sim$ 6.4 kpc in projection before eventually terminating. From this termination point, the NW plume is observed to extend in the jet direction (northwestwards). However, after a distance of $\sim$ 39 kpc from its base, the rapid change in direction is particularly obvious. There are two pronounced bends, at almost 90$^{\circ}$ -- first, southwestwards and then westwards, and the jet appears to bend once more in the northwest direction at $\sim$ 90$^{\circ}$ towards the tail end of the NW plume to form a bell shape. Within the limitations of sensitivity to large-scale structures in our observations, we suggest that this bending trajectory (northwest -- southwest -- northwest) implied by the NW plume could be episodic, stretching over several tens -- hundreds of kpc into the IGM. The rather striking feature here is the apparent asymmetry in morphology of the two plumes. Unlike the NW plume, the SE plume shows only a wiggle pattern downstream. Since the plumes in WAT sources are generally thought to be analogous to smoke from factory chimneys (i.e., light slow-moving structures strongly affected by bulk motions in their environments), these wiggles can be interpreted as the result of strong interaction between the large-scale flow within the plume (which must be light compared with the external medium) and features of the external environment due to either thermal or ram pressure. We attribute the observed asymmetry in physical size and structure of these two large scale components to 2-D projection effects such that the SE plume if projected in a different direction could exhibit similar known prominent bends as the NW plume. The radio source fades into the noise on these images at $\sim$ 171 kpc from the core but the source is more extended, as seen in lower-resolution images.

\begin{figure*}
	\centering
	\includegraphics[width=1.0\textwidth,height=1.0\textheight,keepaspectratio=true]{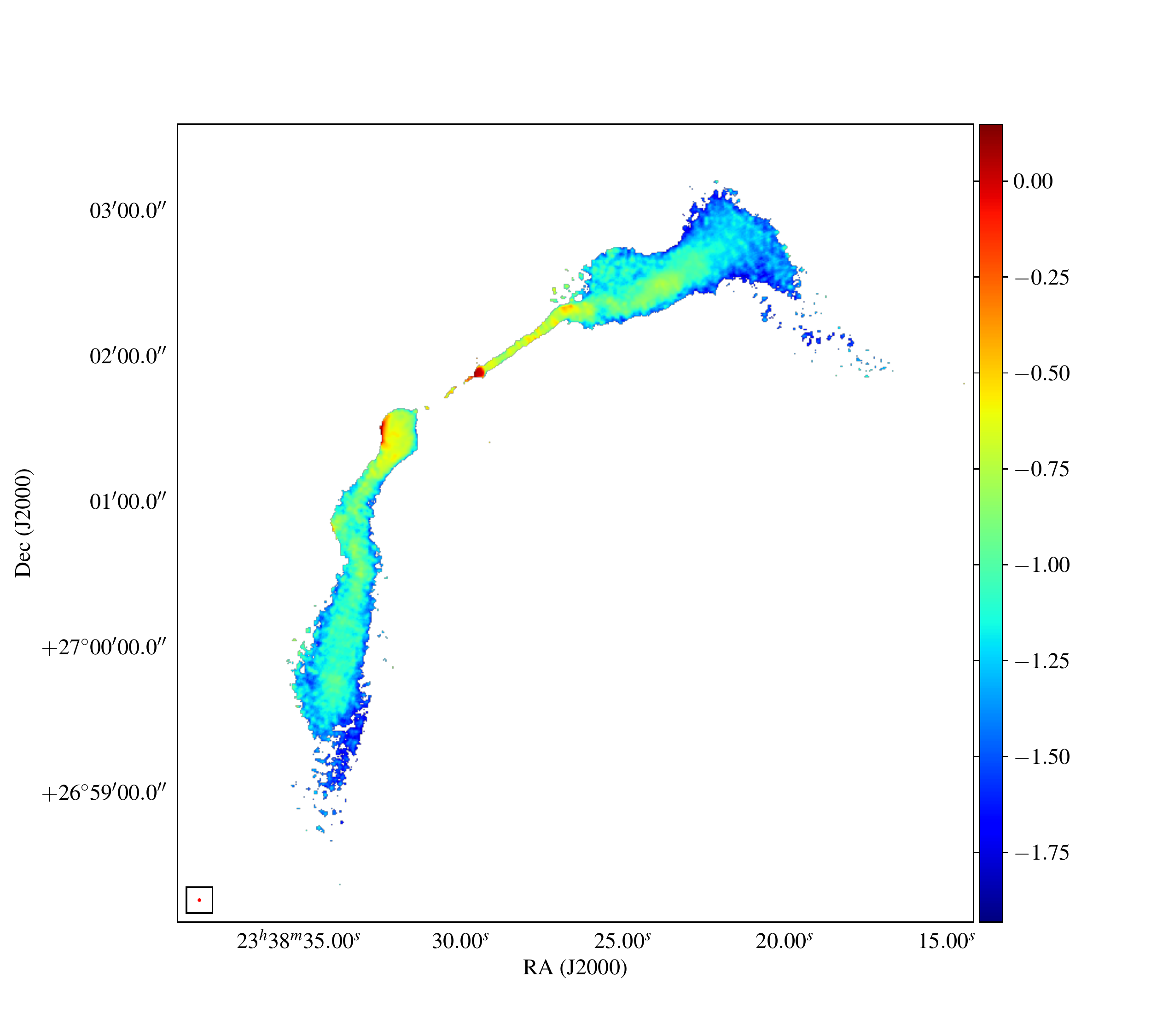}
	\vspace*{-0.75cm}
	\caption{Map of spectral index, $\alpha$ of the WAT radio source 3C\,465 constructed from maps made at two frequencies ($\sim$ 1.5 and $\sim$ 8.5 GHz) at 1.5--arcsec resolution. $\alpha$ is in the range $-$0.5 to $-$0.8 and $-$1.1 to $-$2.3 over the jet and plume regions respectively and is plotted at 3$\sigma$ rms noise cut-off in total intensity.}
	\label{fig:specindxmap}
\end{figure*} 

\begin{figure*}
	\centering
	\includegraphics[width=1.0\textwidth,height=1.0\textheight,keepaspectratio=true]{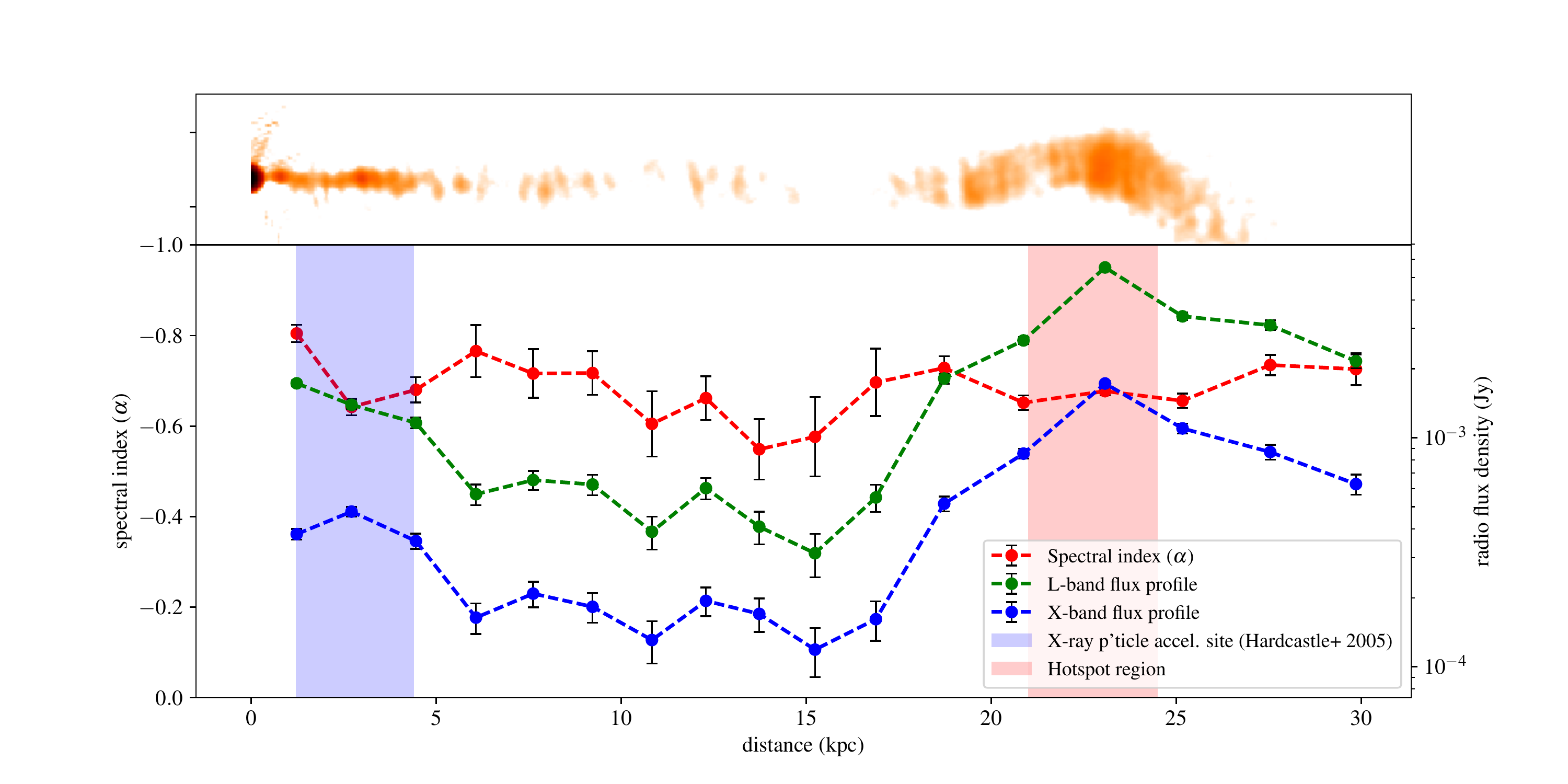}
	\vspace*{-0.75cm}
	\caption{Profile of spectral index, $\alpha$ along the NW jet axis as a function of projected distance from the galaxy central region. Also shown are the respective flux density distributions measured at 1.5 and 8.5 GHz at a resolution of 0.5-arcsec. Shown on the plot are also regions of X-ray enhancements, and so of rapid particle acceleration, as observed by \citet{Hardcastle05+} in a \textit{Chandra} and \textit{XMM-Newton} study of 3C\,465 -- colour coded in blue tone; and the NW hotspot region -- colour coded in red tone. The top panel shows an $\sim$ 45$^{\circ}$ westward rotation of the NW jet used in our spectral analysis.}
	\label{fig:specindxjet}
\end{figure*} 
\subsection{Jet speed and sidedness ratio}
\label{sec:3.4}  
\vspace{-4pt}
The good image fidelity of our maps allows us to estimate the jet/counterjet ratio and place constraints on the jet speed and angle to the line of sight. Following the procedure defined by \citet{Hardcastle98+}, and using our 1.5 GHz VLA map at 1.5-arcsec resolution, we measured only the straight part of the jet and counterjet, over equivalent angular extent -- to avoid any ambiguity in the angle with the line of sight over the integration region. We measure flux densities of 60.16 $\pm$ 0.66 mJy and 4.05 $\pm$ 0.14 mJy for the jet and counterjet respectively, and obtain a sidedness ratio of 14.85 $\pm$ 0.80. Our estimated sidedness ratio is approximately 178 $\%$ higher than that obtained by \citet{Hardcastle&Sakelliou04} for this WAT jet at 8.5 GHz, and is probably a better estimate due to the improved bandwidth capabilities of the new VLA deployed in our observations. We assume that the jets are intrinsically symmetrical and that the observed jet flux asymmetries are due to relativistic beaming effects. This allows us to constrain the characteristic beaming speed, $\beta_{\rm j}$ and angle it makes with our line of sight, $\theta$ by defining the ratio of the jet and counterjet flux densities, $J_{\nu} = S_{\rm j}$/$S_{\rm cj}$ as:
\begin{equation}
J_{\nu} = \left(\frac {1 + \beta_{\rm j} \cos \theta} {1 - \beta_{\rm j} \cos \theta} \right)^\delta 
\label{eqn2}
\end{equation}
where, $\beta_{\rm j}$$c$ is the speed of the jet, which is inclined at an angle, $\theta \in \big[ 0, {\pi}$/${2} \big]$ to our line of sight, and $\delta = m - \alpha$; the constant~ $m = 2$ for a continuous jet (see \citealt{Scheuer&Readhead79} for review). $\alpha$ is the spectral index, which is taken to be $-$0.7. 
\par The observed jet/counterjet asymmetry favours relativistic speeds in the jet -- at least in the regions close to the inner core. From our analysis, we find that $\beta_{\rm j}$$\cos\theta$ = 0.46. Since a lower limit on $\cos\theta$ corresponds to an upper limit on $\theta$, we obtain lower and upper limit values of 0.5 and 61$^{\circ}$ for $\beta_{\rm j}$ and $\theta$ respectively. We find our estimated jet speed, $\nu_{\rm j}$ $\gtrsim$ 0.5$c$ to be consistent with the range of values (0.3 -- 0.7)$c$ obtained by \citet{Jetha06+} in their study of jet speeds in a sample of 30 WAT radio galaxies including 3C\,465. We note here possible velocity stratification -- i.e., it is entirely possible that the centres of the jets have significantly higher speeds but contribute rather little to the observed emission.

\section{Spectral analysis and results}

\subsection{Spectral-index estimates and mapping}
\label{sec:4.1} 

Since our data cover only two frequencies, we have directly calculated the spectral indices as: $\alpha^{\nu_2}_{\nu_1} =$ $\ln$\big[$S$($\nu_1$)/$S$($\nu_2$)\big]\big/$\ln$($\nu_1$/$\nu_2$), by constructing polygonal regions for flux density measurements within the convolved equivalent maps. Where necessary we used polygons slightly wider than regions of \textit{real} emission to reduce sensitivity to any residual misalignments between the maps. Background was taken from the rms errors in total intensity $I$ from circles of radius 10.96 arcsec fixed in position. Since the rms error just depends on the ratio $\nu_1$/$\nu_2$, by standard error propagation we can estimate fractional errors in $\alpha$ across the two frequencies as:

\begin{equation}
\sigma_{\alpha} = \frac{\sigma_{\rm I}\left/{R}\right.} {\ln \left( \frac{\nu_1}{\nu_2} \right)}
\label{eqn3}
\end{equation} 
if ($\sigma_{\rm I}$/$R$) $\ll$ 1 --  where $\sigma_{\rm I}$ is the fractional error on the ratio of the two flux densities and $R$ is the ratio of the two flux densities itself -- the error of which depends on the error (noise levels) on the maps, and $\nu_1$ and $\nu_2$ are frequencies corresponding to the L-band ($\sim$ 1.5 GHz) and X-band ($\sim$ 8.5 GHz) respectively used in our present analysis. 
\par We emphasize that this is a relatively crude model for estimating the error on individual flux density values across the two frequencies. However if we adopt a similar noise level over different regions of flux density integration in our maps and assume that errors in $I$ have a Gaussian distribution with zero mean and rms $\sigma_{\rm I}$ in the image plane and that they are independent on scales larger than the synthesized beam, then our errors should be robust. For consistency, spectral indices at the two resolutions were independently examined at the hotspot region, and this yielded values of $-$0.73 $\pm$ 0.01 and $-$0.75 $\pm$ 0.01 for the 0.5$''$(1.5$''$) resolution respectively, showing that the two sets of maps are consistent. In constructing our spectral-index map we took background to be the off-source noise level $\sigma_{\rm off}$ in $I$ and created the map at 3$\sigma$ cut-off at this rms value over the two frequencies. Finally, due to the complex structure of our source, we have carefully estimated our distances taking into consideration the trajectory of the radio jet to account for the bending of the tail of the radio source through the IGM. All flux profiles are shown on logarithmic scales. Except where explicitly stated, all distances are projected distance measured along the source's ridge line with respect to the central unresolved feature, coincident with the nucleus (core) of the host galaxy. 

\subsection{Spectral index map, distribution and properties}
\label{sec:4.2} 

Figure \ref{fig:specindxmap} above shows a false-colour image of the spectral index for 3C\,465. With the exception of the unresolved core, which is partially optically thick with $\alpha$ $\sim$ $-$0.4, the emission typically has $-$0.5 $\leq$ $\alpha$ $\leq$ $-$0.8 over the jet region and $-$1.1 $\leq$ $\alpha$ $\leq$ $-$2.3 in the plumes. Close to the edges of the jet small errors in deconvolution can cause significant changes in $\alpha$ (i.e., high or low spectral index values), and this is also true for regions where the signal-to-noise ratio is low. The mean spectral index of the jet and counterjet (estimated from our spectral index map) is $-$0.70 $\pm$ 0.04 and, as discussed in section \ref{sec:4.3.1}, there is little spectral variation. The spectral index of the plumes is observed to steepen rapidly with distance from the AGN. This spectral behaviour is as expected in the standard model in which the radio tails flow slowly away from the host galaxy and eventually radiative ageing reduces the number of high-energy electrons resulting in steeper spectra away from the core -- this is similar to the spectral index map of the WAT source 3C\,130 by \citet{Hardcastle98}. 
\par A distinctive feature of our spectral index map (see Fig. \ref{fig:specindxmap}) is the long flat-spectrum component which leaves the NW hotspot and extends in the jet direction (northwest) into the base of the NW plume, possibly suggesting some continued collimated outflow of plasma into the large-scale structures. The component eventually terminates in a broad bright region in the NW plume. A similar feature is seen in the WAT source 3C\,130 (\citealt{Hardcastle98}), and \citet{Hardcastle&Sakelliou04} in a broader WAT sample. Although Hardcastle and Sakelliou did not directly observe our spectral index behaviour, they observed that the jets in some sources of their sample propagate without disruption for some distance into the plumes, and with such additional evidence we suggest that this feature could well be a universal characteristic of this class of radio source. 

\subsection{Spectral profiles}
\label{sec:4.3} 

\subsubsection{The jet and knots} 
\label{sec:4.3.1} 

At 0.5-arcsec resolution, the counterjet in our map is not bright enough to be imaged well and so our analysis here is limited to the NW jet of the source. Individual integrated spectral indices at locations chosen across the jet (using regularly spaced polygons) are plotted in Figure \ref{fig:specindxjet}. These show a fairly constant spectrum over almost the entire jet. Our estimate of $\sim$ 29.8 kpc for the jet length is approximately 6 $\%$ increase to the termination length of the jet in 3C\,465 reported by \citet{Hardcastle&Sakelliou04}. This arises from our inclusion of the trajectory in measuring distances along the jet to account for the bends in the radio source. The large $\alpha$ errors from $\sim$ 10--17 kpc correspond to regions of low surface brightness of the radio jet. 
\par The comparatively constant spectral index of the jet spans $-$0.6 $\leq$ $\alpha_{\rm jet}$ $\leq$ $-$0.8 with an average value of <$\alpha_{\rm jet}$> = $-$0.70 $\pm$ 0.04. As shown in Figure \ref{fig:specindxjet}, within the first 4.45 kpc of radius from the core, the spectral indices flatten from $-$0.80 $\pm$ 0.02 to $-$0.64 $\pm$ 0.02 ($\Delta \alpha$ = $-$0.16) and thereafter increase marginally to $-$0.68 $\pm$ 0.03 ($\Delta \alpha$ = $-$0.04). The spectral flattening over this geometric area indicates the presence of a young electron population and provides evidence of ongoing particle acceleration at the jet base. This is consistent with X-ray evidence that particle acceleration is found at the base of the radio jet within the knotty structures (\citealt{Hardcastle05+}). The rather interesting observation here is that the region of flattest spectral index ($\alpha$ = $-$0.55 $\pm$ 0.07) at a distance of $\sim$ 13.7 kpc does not correspond to either of the bright knots at the jet base or the compact region of intense radio emission (hotspot) nor to any feature in the X-ray data. Beyond 13.7 kpc, there is systematic steepening of the spectra from $-$0.55 $\pm$ 0.07 to $-$0.73 $\pm$ 0.04 until the bright compact region (hotspot). The spectrum over three measurements in the region of intense radio emission is almost constant, with <$\alpha$> = $-$0.66 $\pm$ 0.01. 

\subsubsection{The hotspots} 
\label{sec:4.3.2} 
Integrated spectral profiles across the two hotspot (NW and SE) regions separately are plotted in Figure \ref{fig:specindxhotspot}. For reliability and purposes of direct comparison, we used regularly spaced polygons across a circle of diameter 16.05 arcsec to estimate the hotspot spectra. Due to the relatively smooth transition of the NW jet into the base of the NW plume, our estimate for the NW hotspot may be overestimated by a few tens of per cent in our present analysis and the opposite is true for the SE hotspot region which features a comparatively broad bright structure. The spectral indices over the integration regions in the hotspots show little difference between the two components with mean values of <$\alpha_{\rm NWh}$> = $-$0.69 $\pm$ 0.01 and <$\alpha_{\rm SEh}$> = $-$0.65 $\pm$ 0.01. This plausibly indicates that an electron population with the same properties is injected at these sites -- which lie at similar distances ($\sim$ 23 kpc) on either side of the host galaxy. 
\par The NW hotspot spectrum (Fig. \ref{fig:specindxhotspot} upper panel) flattens from $-$0.73 $\pm$ 0.01 to $-$0.61 $\pm$ 0.01 within 19--21 kpc. Beyond this, the spectrum steepens to $-$0.77 $\pm$ 0.01 and thereafter flattens again to $-$0.62 $\pm$ 0.01 at $\sim$ 24.5 kpc, before experiencing a further steepening beyond 25 kpc. A significant tendency for the spectral index to flatten over the SE hotspot is apparent from Figure \ref{fig:specindxhotspot} (lower panel). The effect is subtle ($\Delta \alpha \leq$ $-$0.05) nonetheless consistent; with a steady decline in $\alpha$ from $-$0.76 $\pm$ 0.01 to $-$0.57 $\pm$ 0.01 ($\Delta \alpha$ = $-$0.2), plausibly indicating acceleration of high energy electrons as would be expected in the case of typical classical double (FR II ) radio galaxies. Unlike the SE hotspot which show a steady flattening with increasing distance from the core, there is no clear single trend in the NW hotspot spectra; this is expected from the larger random errors in the NW hotspot due to its comparatively narrower distribution on the sky. 
\par Indeed from the standard model, if the spectrum is not a pure power-law and the jets are relativistic, then systematic variation between the spectra of the jet and counterjet is expected. Overall, the SE hotspot has a spectral index flatter than the NW hotspot, and we find little variation between the spectral profiles of the two components over our geometric area of integration in the present analysis. Since these compact regions of intense radio emission, at least in FR IIs, are sites of AGN jet termination with consequent interaction with the lobe material (or plume material in the case of FR Is) forming strong shocks to yield physical conditions required for particle acceleration (e.g. \citealt{Massaglia07}), it is quite enticing to associate spectral flattening in the hotspots of our sample (which falls at the FR I/FR II break) with such sites of high energy particle acceleration. This is particularly the case for the SE hotspot which has a flatter spectrum than the NW jet and hotspot over almost the entire integration region in the present analysis. 

\subsubsection{The plumes} 
\label{sec:4.3.3} 
\begin{landscape}
	\begin{figure}
		\begin{multicols}{2}
			\includegraphics [scale=0.67]{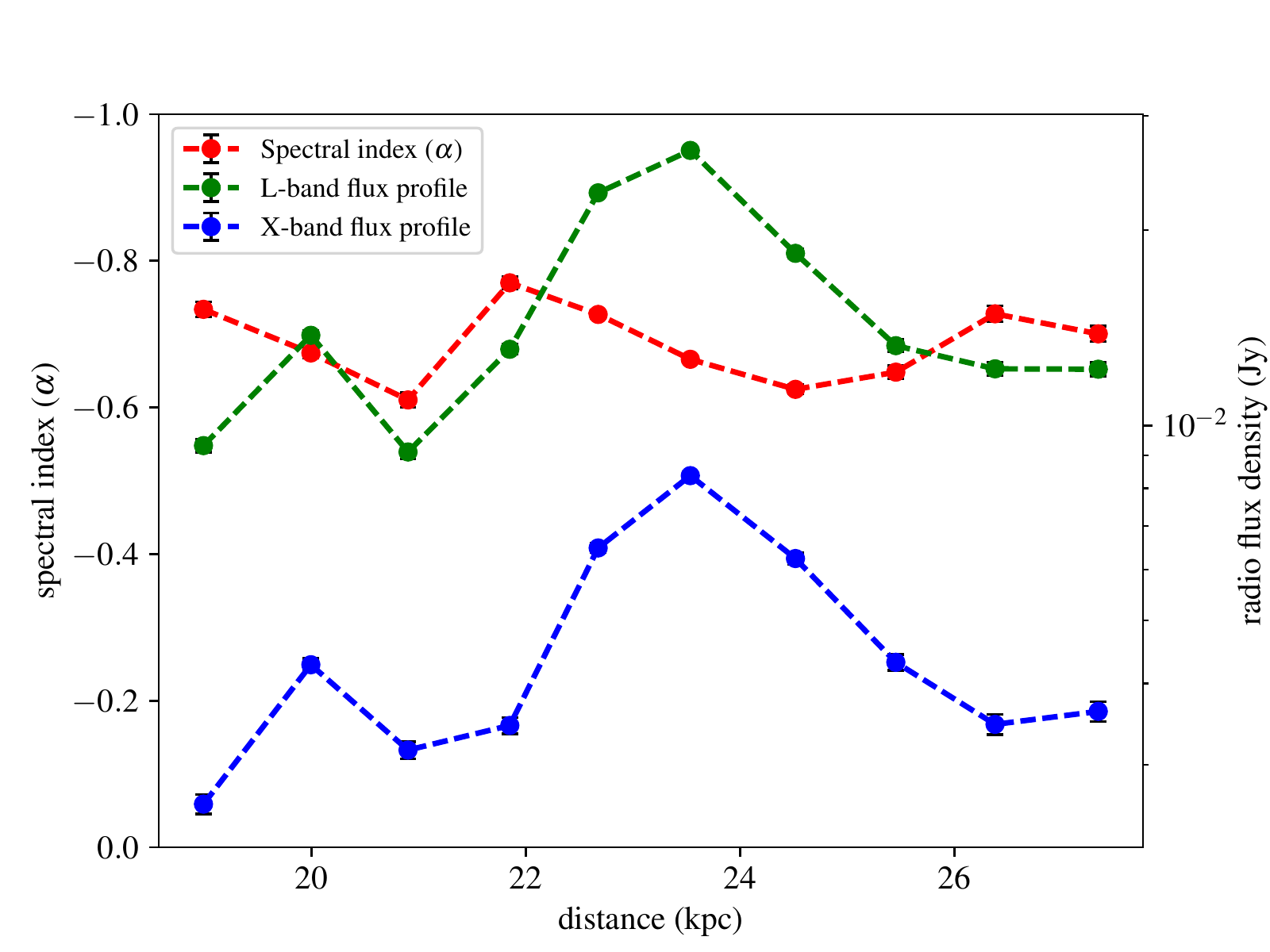}
			\includegraphics [scale=0.67]{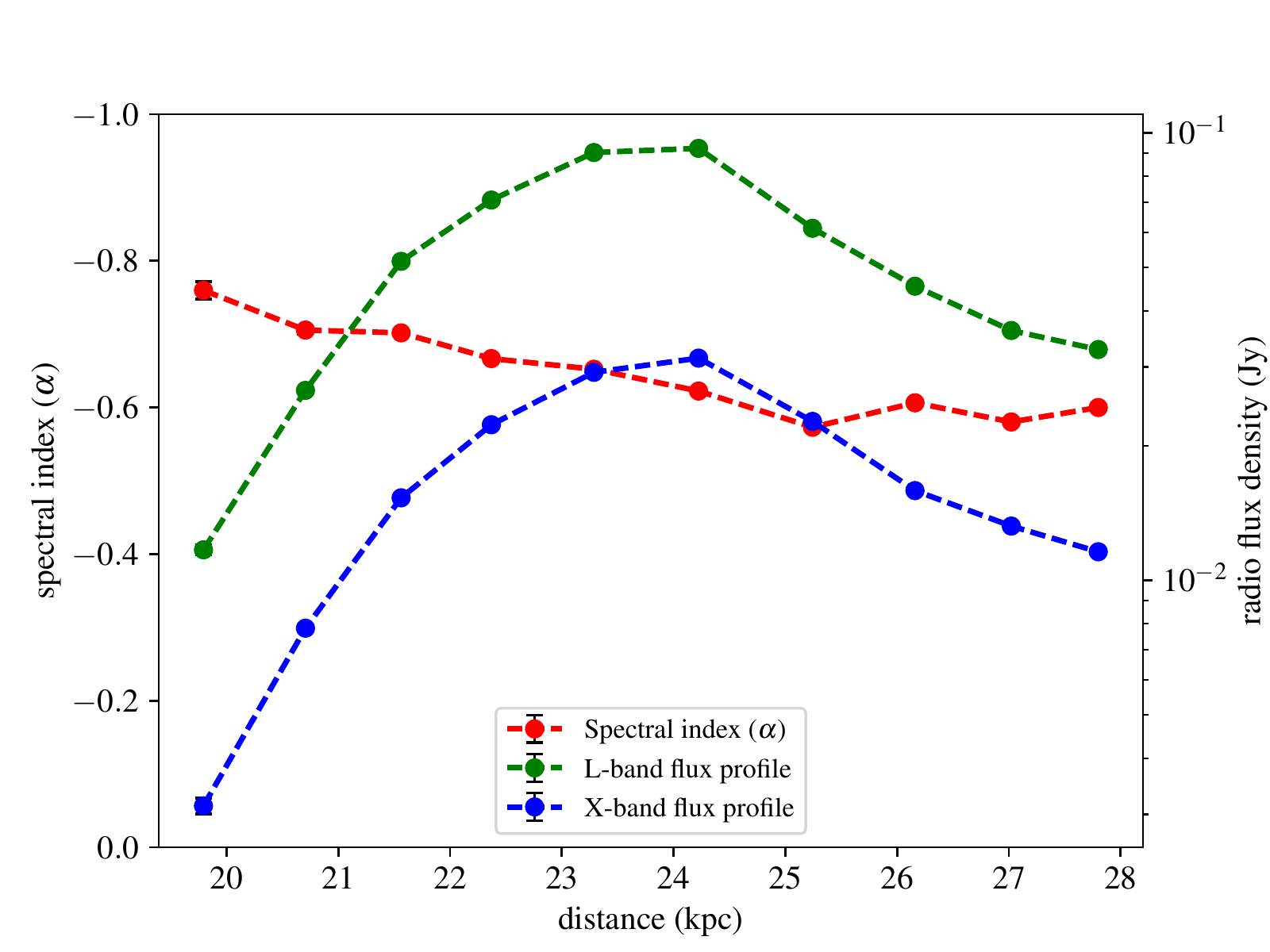}
			\vspace*{-0.15cm}
			\caption{Distribution of spectral indices, $\alpha^{8.5}_{1.5}$  for the two hotspot regions plotted as a function of distance from the nucleus. Also shown are the respective flux density distributions. Upper panel: NW hotspot. Lower panel: SE hotspot. The spectral indices were calculated from total intensity using regularly spaced polygons across a circle of diameter 16.05 arcsec which was centred on the two components; and error bars as in (Fig. \ref{fig:specindxjet}).}
			\label{fig:specindxhotspot}

			\includegraphics [scale=0.67]{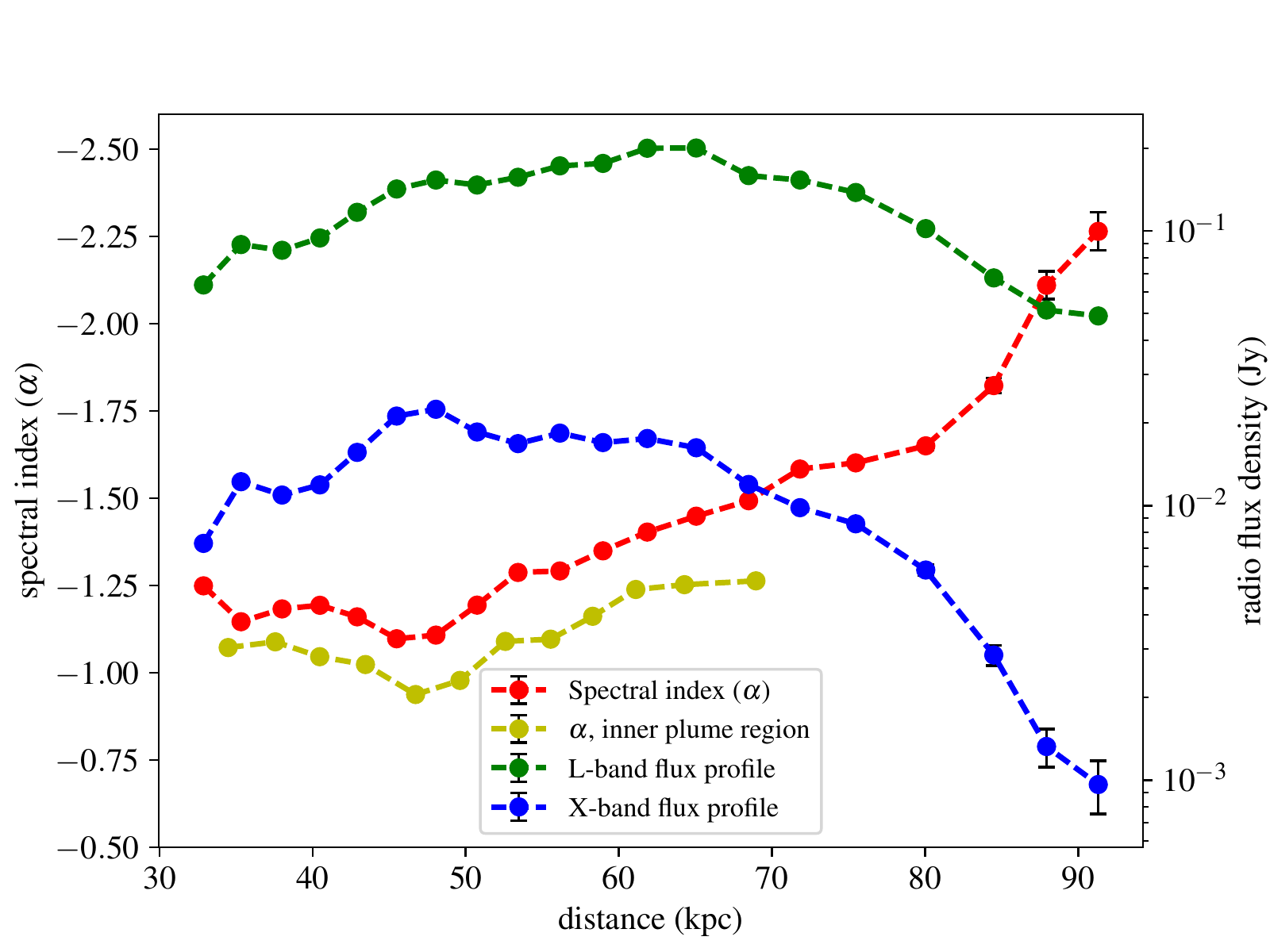} 
			\includegraphics [scale=0.67]{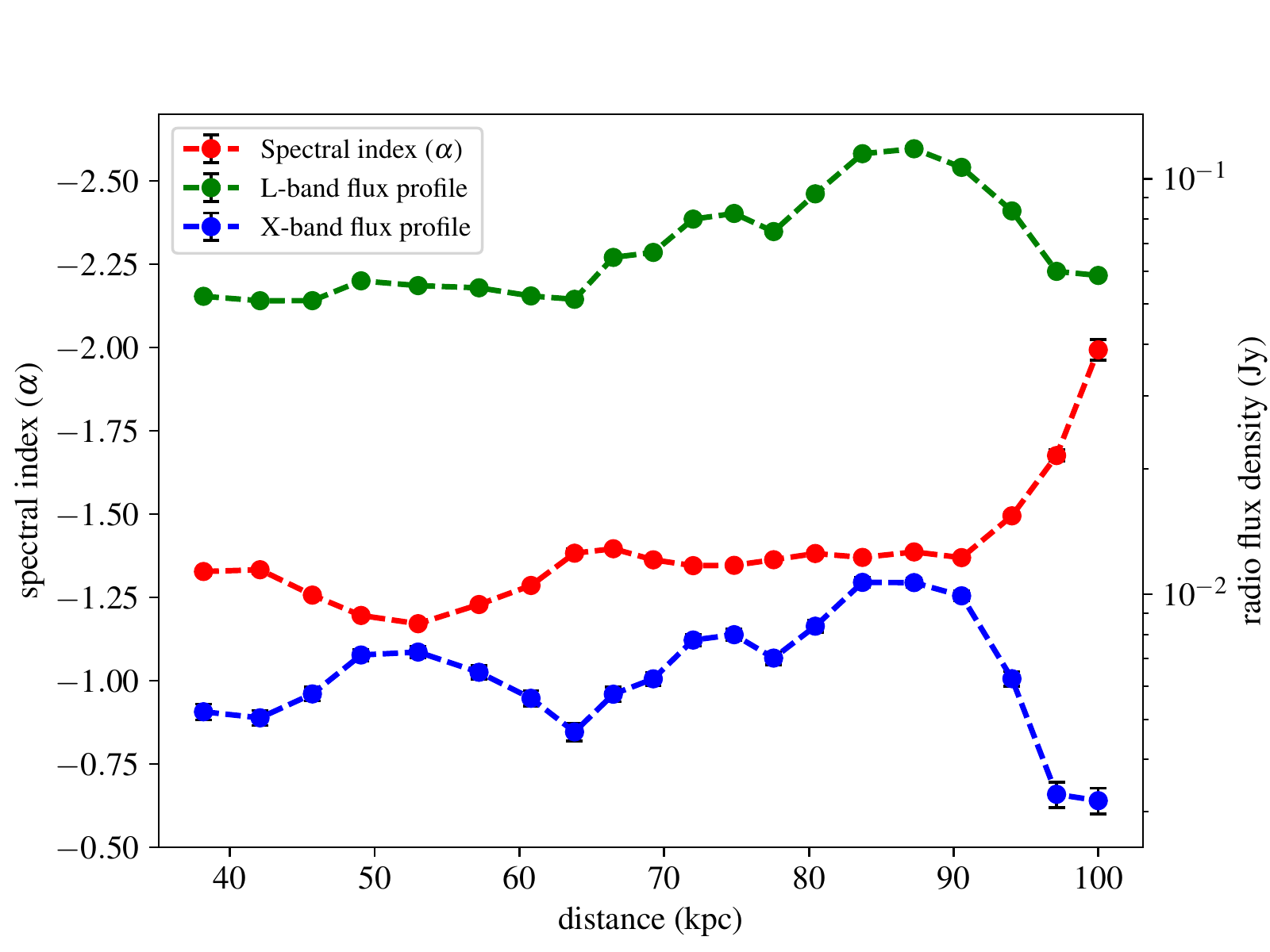}
			\vspace*{-0.15cm}
			\caption{Variation in spectral indices, $\alpha^{8.5}_{1.5}$ with fiducial distance in the extended structures of 3C\,465. Upper panel: NW plume; also shown here is a profile of the continued collimated outflow at the base of the NW plume as seen in total intensity (Fig. \ref{fig:emlnvlamap}). Lower panel: SE plume. Estimate of $\alpha$ and corresponding errors are as quoted in (Fig. \ref{fig:specindxjet}).}
			\label{fig:specindxplume}
		\end{multicols}
	\end{figure}
\end{landscape}

We used 20 polygon slices each along the plumes for estimating the spectral indices in the large-scale components. The profiles are plotted in Figure \ref{fig:specindxplume} with their corresponding flux densities and errors. These show a steady steepening of $\alpha$ in the NW plume compared with the practically constant $\alpha$ for the SE over the majority of the region. However, towards the tails of both plumes there is a rapid steepening in the spectral index from $-$1.60 $\pm$ 0.01 to $-$2.26 $\pm$ 0.05, and $-$1.37 $\pm$ 0.01 to $-$1.99 $\pm$ 0.03 for the NW and SE plumes respectively. As our model (see section \ref{sec:4.4}) suggests, the steepening is consistent with synchrotron theory; and except for the final 3 locations at the tail end of each spectrum, there is no significant dispersion (<$\delta \alpha$> = $\pm$ 0.01) between the SE and NW plumes. The mean difference between the NW and SE plume is <$\alpha_{\rm NWp}$ -- $\alpha_{\rm SEp}$> = $-$0.05 $\pm$ 0.02. By this measure, the spectral steepening between the two large-scale components is not very significant; consistent with a single spectral index change for both plumes. 
\vspace{-50pt}
\par For reasons discussed in section \ref{sec:4.2} above, we have examined the spectral gradients of the NW plume in more detail. We have plotted profiles of spectral index in the plumes in two representations: as unaveraged slices across -- (i) the whole plume, and (ii) inner region (out to $\sim$ 61.5 arcsec) to show the level of variation on large and small scales respectively; and to compare $\alpha$ in this region with the overall plume. The profile is as shown in Figure \ref{fig:specindxplume} (upper panel). The clear tendency for the spectra to flatten slightly within the inner regions of the NW plume is confirmed; with the brightest spot in the region coinciding with comparatively the flattest spectra ($\alpha$ = $-$0.94 $\pm$ 0.09) -- indicating the injection of young electron population into the base of the NW plume, and possible acceleration of particles at this site. 
\par In general the spectrum of the NW plume is steeper <$\alpha_{\rm NWp}$> = $-$1.43 $\pm$ 0.01 compared with the SE <$\alpha_{\rm SEp}$> = $-$1.38 $\pm$ 0.01, and this is likely the consequence of variations in mass injection and external pressure and density gradients in the two regions. Conventionally, as the radio jet breaks through the dense ICM and transitions into plumes the spectra will steepen further away from the AGN due to radiative ageing and expansion effects; and this is consistent with our observed spectral behaviour in the two components. Figure \ref{fig:specindxplume} shows a systematic pattern in the spectral distribution measured in both plumes, from which we infer that similar flow dynamics are at play in the two components, with the asymmetry in total intensity a consequence of projection effects and flow differences. Additional evidence for this conclusion can be drawn from the good agreement between the integrated spectral indices measured at the initial 8 locations of the SE plume and the spectral profile of the inner regions of the NW plume; both show a similar trend. Generally the spectrum is fairly constant in both plumes for the first 70 kpc in the NW and 90 kpc in SE, after which they steepen. 
\par Overall our spectral profiles suggest that the plumes are approximately homologous structures, in the sense that there is a clear trend in their spectral distributions despite evidence of considerable local variations in physical size and structure. While the uncertainties are large near the plume tails, there is evidence for a mild trend toward a steeper spectral indices with decreasing flux density in both plumes. We note that the spectral index limits originate from the limiting flux density of the least-sensitive frequency ($\sim$ 8.5 GHz) in our two-point spectral index calculation.
%%%
%%%
%%%

 \subsection{A model for the synchrotron loss time} 
 \label{sec:4.4} 
 The electron distribution arising from shock acceleration of a synchrotron emitting population of relativistic plasma is assumed to be a broken power law in momentum: $n$($p$) $\propto p^{-s}$, $p_{\rm min}$ $<$ $p$ $<$ $p_{\rm c}$; and $n$($p$) $\propto p^{-\left(s+\Delta s\right)}$, $p_{\rm c}$ $<$ $p$; where $s$ is governed by the injection process, and $\Delta s$ = 1 for synchrotron ageing with ongoing injection in a constant $B$ field (see \citealt{Eilek96} and references therein). Since the synchrotron and inverse-Compton fractional energy loss rates are proportional to gamma, this injection spectrum develops a break where the losses become important. The break shifts to lower energies over time, so that the break frequency would be expected to decrease over time, in the absence of strong reacceleration of the electrons. The synchrotron lifetime, $\tau_{\rm syn}$ of the electron population as a whole can be extracted from the break frequency, $\nu_{\rm br}$, and the observed radiation spectrum in a given frequency range will be observed to steepen as $\nu_{\rm br}$ passes through that frequency range. Here, we set some limits on the source properties from a simple kinematic model for the synchrotron lifetime of the radio tails in 3C\,465.
\par
\begin{figure*}
	\centering
	\includegraphics[width=0.9\textwidth,height=0.9\textheight,keepaspectratio=true]{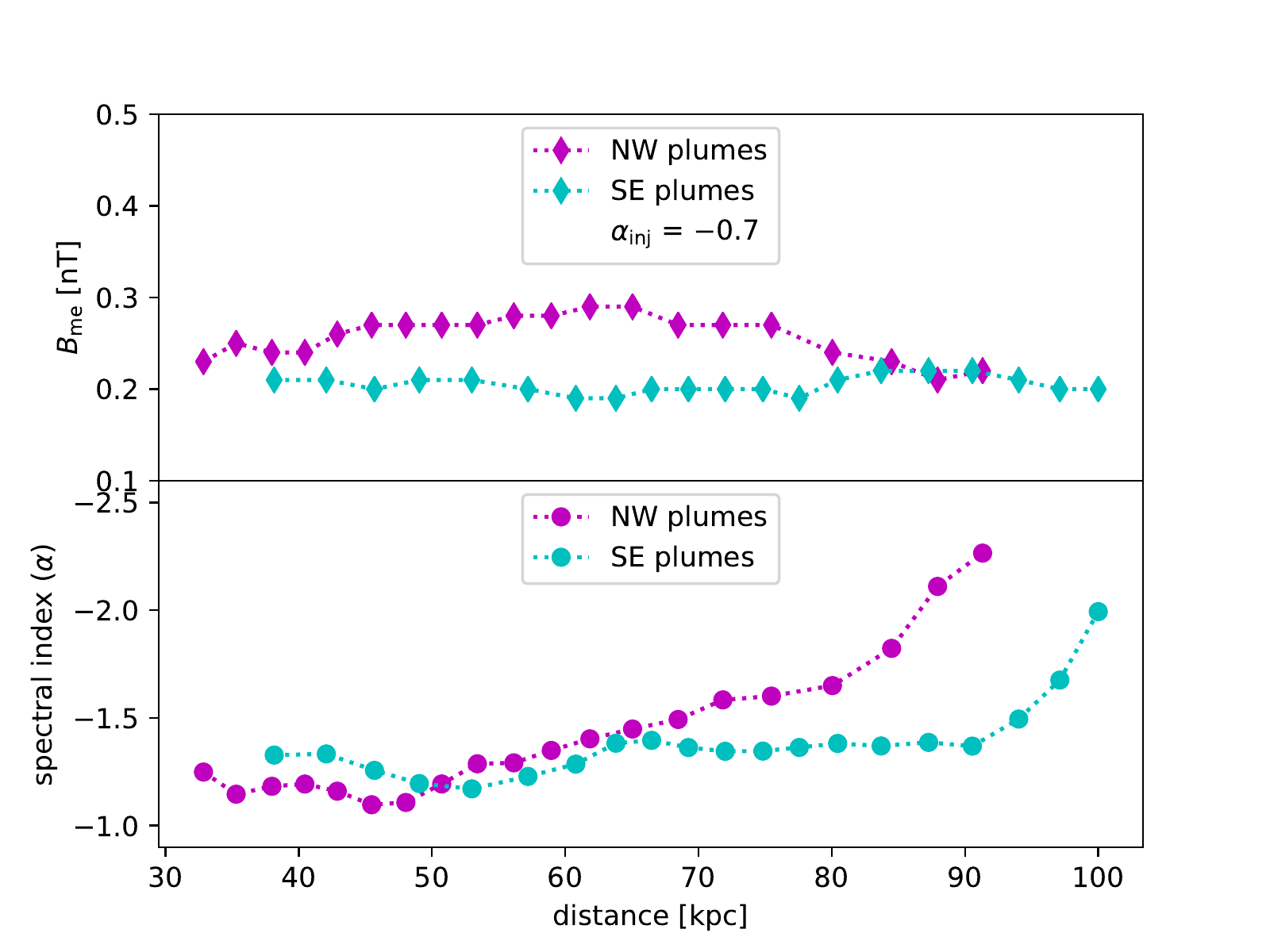}
	\vspace*{-0.45cm}
	\caption{How the minimum-energy field strength, $B_{\rm me}$ (upper panel) compares with our estimated point-to-point spectral index, $\alpha^{8.5}_{1.5}$ (lower panel) evolution in the plumes of 3C\,465.}	
	\label{fig:dynage}
\end{figure*}

 Since the synchrotron spectrum of an emitting relativistic plasma depends on both electron energy and field strength, a knowledge of the field strength, together with the spectral index and break frequency are essential parameters for estimating the flow timescale of the synchrotron plasma. To progress, we assume minimum energy conditions, and estimate the magnetic field strengths for the radio tails in 3C\,465 using the formalism of \citet{Worrall&Birkinshaw06}, which is an analytical treatment of the minimum-energy magnetic field.
%%%
%%%
\begin{equation}
B_{\rm me} = \left[ \frac {\left(\alpha + 1 \right) C_1}{2C_2} \frac {\left(1 + \kappa \right)}{\eta V} L_{\nu} \nu^{\alpha} \frac {\left(\gamma_{\rm max}^{1 - 2\alpha} - \gamma_{\rm min}^{1 - 2\alpha} \right)}{\left(1 - 2\alpha\right)}	\right]^{1 \left/\right. \left(\alpha + 3 \right)} 
\label{eqn4}
\end{equation}
where $C_{\rm 1}$ and $C_{\rm 2}$ are constants that depend on $\alpha$ and field inclination, $L_{\nu}$ is the spectral luminosity in the emission region, of volume, $V$, at frequency, $\nu$. The parameters $\kappa$, $\gamma_{\rm min}$ , $\gamma_{\rm max}$ and $\eta$ are defined in Table \ref{tab:table3} where their adopted values are also given. The field strength (here taken as the minimum energy value, $B_{\rm me}$) together with $\nu_{\rm br}$, then allow us to determine the radiative lifetime of the electrons downstream of the tails from:

\begin{equation}
\tau_{\rm syn} = \frac {1.6\times10\,^{4} \,\, B\,^{0.5}}{(U_{\rm B}\, + \, U_{\rm CMB})} \left[ (\nu_{\rm br}\left(1+z\right) \right]^{-0.5} \,~\,~\,~\,~\,~{\rm yr}		 
\label{eqn5}
\end{equation}
where $B$ (in tesla) is the magnetic field strength,\, $U_{\rm B}$ (J\,m$^{-3}$) is the energy density in the magnetic field,\, $U_{\rm CMB}$ = 4.2$\times$10$^{-14}\left(1 + z\right)^4$\, J\,m$^{-3}$\, is the equivalent energy density of the cosmic microwave background radiation at redshift\, $z$,\, and\, $\nu_{\rm br}$\, is expressed in units of Hz. Equation (\ref{eqn5}) accounts for both synchrotron and inverse Compton losses. Since our spectral indices were derived from two-frequency data, we chose 1.5 GHz as the upper limit on the break frequency for the steep-spectrum plumes, which puts a lower limit on the synchrotron lifetime scale. For flat radio spectra ($\alpha$ $<$ $-$0.6) and steep radio spectra ($\alpha$ $>$ $-$0.7) classical estimates of the minimum energy field strengths are respectively undersetimated and overestimated by a few per cent (See \citealt{Beck&Krause05} and references therein).  Thus, our estimated mean spectral index of\, $\sim$\, $-$1.4\, for the NW and SE plumes is likely to be about twice the spectral index of the unaged synchrotron spectra for which we adopt $-$0.7 as the injection spectral index (see Table \ref{tab:table3}) in our minimum energy calculations. We used the spectral index, volume, and luminosity for each of the 20 polygonal regions to estimate the minimum-energy values and the corresponding $\tau_{\rm syn}$ along the radio tails from Equations (\ref{eqn4}) \& (\ref{eqn5}). The minimum energy values are listed in Table \ref{tab:table4}, and the field strength\, $B_{\rm me}$\, together with our derived point-to-point spectral indices ($\alpha^{8.5}_{1.5}$) as a function of projected distance down the plumes are plotted in Figure \ref{fig:dynage}. 
\par We find that our estimated minimum-energy field strengths range from 0.21 -- 0.29 nT in the NW plume, and 0.19 -- 0.22 nT in the SE plume. The mean field energy density is marginally higher in the NW plume (<$U_{\rm B}$> = 0.03 J\,m$^{-3}$) compared with the SE plume (<$U_{\rm B}$> = 0.02 J\,m$^{-3}$). This could arise from a systematic difference in projection effects (and hence volume calculations) on the two sides. The estimates of radiative lifetime of electrons emitting at $\nu_{\rm br}$ = 1.5 GHz in the two plumes at all distances are very similar, at about 90 Myr.
\par In typical FR Is, and under the so called {\textquotedblleft{velocity expansion}\textquotedblright} assumptions (e.g. \citealt{Giacintucci08+}), the older electron emitting populations must exist at larger distances from the core since there is a constant injection of a fresh population of electrons near the core. However, the spectral steepening we observe is inconsistent with the ageing of the plasma towards the tail ends in both plumes. Our model indicates that the radiative lifetime is relatively constant with downstream distance in both plumes. This implies that new flat-spectrum electrons are produced at the base of the plume materials. The presence of these young electron populations at such large distances away from the AGN supports a working picture in which there is a continuous reacceleration of particles along the radio tails of 3C\,465; the diffusion time $\tau_{\rm D}$ for electrons to travel some tens of kpc must therefore be longer than our estimate of 90 Myr for $\tau_{\rm syn}$. This assumption is reasonable for 3C\,465 since typical diffusion timescales are of the order of a few $\sim$ 10$^2$ Gyr (e.g. \citealt{Berezinsky97+}). This in principle requires that acceleration must be more efficient in the outer parts of the source than in the jets; and supports a continuous injection of relativistic plasma from the hotspots into the base of the plume materials as observed in total intensity and from our spectral index map. We note here that $\tau_{\rm syn}$ is an estimate of the lifetime to energy losses for electrons radiating at frequency, $\nu_{\rm br}$. There may also be losses from adiabatic expansion and/or reacceleration with some different properties (which is not accounted for in our model) and, therefore, the radio source must be older than our inferred apparent synchrotron lifetime.

\begin{table}
	\begin{threeparttable}
		\caption{Assumed values of the model free parameters}
		\label{tab:table3}
		\begin{tabular}{|l|c|c|c|r|} 	
			\hline
			{Parameter} &\multicolumn{1}{c}{Symbol} &\multicolumn{1}{c}{Value}\\
			(1) &(2) &(3)\\ 
			\hline
			{Min. electron Lorentz factor (injected)} &{$\gamma_{\rm min}$} &{10}\\
			{Max. electron Lorentz factor (injected)} &{$\gamma_{\rm max}$} &{10$^{7}$}\\
			{Energy ratio of non-radiating particles} &{$\kappa$} &{0}\\
			{Filling factor} &{$\eta$} &{1}\\
			{Injection spectral index} &{$\alpha_{\rm inj}$} &{$-$0.7}\\
			\hline
		\end{tabular}
		\begin{tablenotes}
			\small \item{{$\textbf{*}$} Our model assumes cylindrical geometry of the emission region.}
		\end{tablenotes}
	\end{threeparttable}
\end{table}

\subsection{Implications for particle acceleration in 3C\,465} 
\label{sec:4.5} 

The evolution of radio jet spectra is governed by the balance of particle acceleration and energy-loss processes (e.g. \citealt{Eilek&Shore89}). For a power law distribution -- i.e., an ensemble of homogeneous and isotropic injected population of electrons, $N\left(\epsilon, 0\right)d\epsilon = N_{\circ} d\epsilon^{-\delta}$; there exists a corresponding power law synchrotron spectrum, $S$($\nu$) $\propto \nu^{\alpha}$ with $\delta$ = 2$\alpha + 1$, the logarithmic slope of which relates the Mach number of the accelerating shock. For a detailed account on the subject, see \citet{Hardcastle13} and references therein. A model in which relativistic electrons suffer radiative and adiabatic losses as the radio jet propagates is consistent with the spectral steepening we observe, away from the AGN. We note that relatively high frequency ($\sim$ 8.5 GHz) emission observed at large distances from the hotspots implies plausible \textit{in situ} acceleration of relativistic particles in \textit{at least} some parts of the extended structures. This again agrees with the observed spectral flattening at a few discrete locations in both plumes as seen in our spectral profile plots. 
\par A good assessment of possible jet particle acceleration mechanisms is provided elsewhere (e.g. \citealt{Heavens84}; \citealt{Bisnovatyi&Lovelace95}; \citealt{Rieger07+}; \citealt{Summerlin&Baring12}; \citealt{Liu17+}). Fermi acceleration is the most popular scheme; with first-order Fermi acceleration thought to be the main particle acceleration mechanism at collisionless MHD shocks (e.g. \citealt{Schmitz02+}). Other acceleration mechanisms, the so called {\textquotedblleft second order\textquotedblright} processes are generally thought to be less efficient, but as pointed out by \citet{Ostrowski&Schlickeiser93}, the relativistic particle energy distribution in all cases approaches a power law that reflects the acceleration physics at lower frequencies. It has long been assumed that the particle acceleration process that sets the \textit{energy spectrum} does not occur in the \textit{extended lobes} (plumes) of extragalactic radio sources (e.g. \citealt{Hughes80}). Plumes are normally assumed to be just passive outflows, and although \textit{in situ} wave generation within the plume material may allow particle acceleration there, the fact that we observe the spectra to steepen monotonically with distance from the hotspots does not provide any such evidence. This notwithstanding, in the present study, we posit that acceleration if occurring at all would be more likely distributed throughout the plumes of 3C\,465, rather than restricted to a few localised sites containing peak brightness, and that the observed variation in spectral profiles results from the combined effects of synchrotron and adiabatic (energy transferred to surrounding medium only as work) losses, in addition to the underlying acceleration mechanism, as per discussions below.

\begin{enumerate}
	\item[i.] {Non-relativistic shocks}\\ 
	\indent Contrary to earlier work (e.g. \citealt{Bell78}), later work by \citet{Blanford&Eichler87} in test particle diffusive shock acceleration (test particle approximation) has established that steeper spectra can in principle be produced in weaker, non-relativistic shocks. Since our velocity estimate implies that the jet in 3C\,465 is mildly relativistic at the base, it is appropriate to assume non-relativistic flows/speeds in the extended structures due to deceleration resulting from entrainment by ambient material. Thus, the observed spectral steepening in the plumes is more likely the result of the particle acceleration being by weak, non-relativistic shocks in the plume material. We note that by this conclusion, we are by no means neglecting the likelihood of synchrotron losses as possible alternative explanation for our observed spectra steepening.
	\newline
	\item[ii.] {Mildly relativistic shocks}\\ 
	\indent Previous studies (e.g. \citealt{Summerlin&Baring12}) have shown that, depending on the nature of scattering, shock speed and field obliquity, mildly relativistic shocks can generate a wide range of power law slopes for the energy spectrum. Our observations indicate mildly relativistic speeds in the jet (at least in the inner regions close to the core) and thus would generate energy indices, $\delta$, hence spectral indices that depend on mean flow speed. Mildly relativistic speed upstream of the shock front seems the most likely scheme which meets this requirement. Thus, we argue that the observed synchrotron spectra in the jet of 3C\,465, and more so the spectral flattening in the inner regions close to the core (where the bright knots form), result from mildly relativistic shocks, consistent with the \citet{Laing&Bridle13} suggestion that mildly relativistic shocks are responsible for the range of physical conditions in FR I jet bases.
	\newline 
	\item[iii.] {Ultra-relativistic shocks}\\
	\indent Numerical (e.g. \citealt{Ellison&Double04}) and analytical (e.g. \citealt{Kirk00+}) results have shown that ultra-relativistic shocks in principle can produce a power law energy spectrum with $\delta$ = 2.23 ($\alpha$ = $-$0.62). Comparison with our observed mean $\alpha$ values of $-$0.69 $\pm$ 0.01 and $-$0.65 $\pm$ 0.01 for the NW and SE hotspots respectively suggests that both components have marginally but significantly steeper spectral indices than predictions from theory. Given this discrepancy and our constraints on the jet speed ($\beta_{\rm j}$ $\gtrsim$ 0.5), it seems rather unlikely that ultra-relativistic shocks are relevant to the physical conditions of these compact regions of intense radio emission. However, in the absence of independent statistical evidence, it is plausible that the flow at the hotspots is supersonic everywhere, suggesting strong {\textquotedblleft \textit{relativistic}\textquotedblright} shocks in this flow regime that can be modified by back-pressure of the accelerated particles. This in principle would alter the shape of the synchrotron spectrum as seen in our respective hotspot spectral profiles. 
\end{enumerate}

\begin{table*}
	\begin{threeparttable}
		\caption{Fitted minimum energy values of physical parameters for representative regions along the radio tails (plumes) of 3C\,465}
		\label{tab:table4}
		\begin{tabular}{|l|c|c|c|c|c|c|c|c|c|c|c|c|r|} 
			\hline
			\multicolumn{6}{c}{NW plumes}
			&&&\multicolumn{6}{c}{SE plumes}\\
			\multicolumn{1}{c}{Dist.}&
			\multicolumn{1}{c}{Vol.}&
			\multicolumn{1}{c}{log $L_{\nu}$}&
			\multicolumn{1}{c}{$B_{\rm eq}$}&
			\multicolumn{1}{c}{$U_{\rm B}$}&
			\multicolumn{1}{c}{$\tau_{\rm syn}$}&&&
			\multicolumn{1}{c}{Dist.}&
			\multicolumn{1}{c}{Vol.}&
			\multicolumn{1}{c}{log $L_{\nu}$}&
			\multicolumn{1}{c}{$B_{\rm eq}$}&
			\multicolumn{1}{c}{$U_{\rm B}$}&
			\multicolumn{1}{c}{$\tau_{\rm syn}$}\\
			
			(kpc) &(kpc${^3}$) &(W Hz$^{-1}$) &(nT) &(pJ m$^{-3}$) &(Myr) &&& (kpc) &(kpc${^3}$) &(W Hz$^{-1}$) &(nT) &(pJ m$^{-3}$) &(Myr)\\
			(1) &(2) &(3) &(4) &(5) &(6) &&& (7) &(8) &(9) &(10) &(11) &(12) \\ 
			\hline
			32.9 &209.4 &38.9 &0.24 &0.02 &92 &&& 38.2 &260.7 &39.0 &0.22 &0.02 &93\\
			40.5 &271.0 &39.7 &0.25 &0.02 &91 &&& 49.1 &317.4 &39.6 &0.22 &0.02 &93\\
			50.7 &303.5 &40.6 &0.28 &0.03 &89 &&& 63.8 &374.4 &40.0 &0.20 &0.01 &93\\
			61.9 &330.2 &41.3 &0.30 &0.03 &88 &&& 74.8 &491.2 &40.8 &0.21 &0.02 &93\\
			71.8 &317.3 &41.3 &0.28 &0.03 &90 &&& 83.7 &509.9 &41.4 &0.23 &0.02 &93\\
			80.0 &295.8 &41.1 &0.26 &0.02 &91 &&& 90.5 &461.5 &41.4 &0.23 &0.02 &93\\
			91.3 &217.1 &40.7 &0.23 &0.02 &93 &&& 99.9 &346.5 &41.0 &0.21 &0.02 &93\\
			\hline
		\end{tabular}
		\begin{tablenotes}
			\small \item{Notes: Columns (1), (2), (7) \& (8) are self explanatory. Columns (3) \& (9) -- log of Luminosity at 1.5 GHz; Columns (4) \& (10) -- Equipartion magnetic field; Columns (5) \& (11) -- Minimum energy density in the magnetic field; Columns (6) \& (12) -- Radiative lifetime of synchrotron emitting electrons.}
		\end{tablenotes}
	\end{threeparttable}
\end{table*}

\subsection{Acceleration mechanisms} 

Our analysis shows that first-order Fermi processes at mildly relativistic and non-relativistic shocks are the most probable acceleration mechanism at play in 3C\,465. \citet{Laing&Bridle13} suggest that a likely complication may exist, in the sense that electron acceleration that can produce X-rays requires a distributed system of shocks covering a significant distance along the jet axis rather than restricted to a few localized shock sites. In the present study, such shock systems can be associated with the multiple bright knotty structures at the jet base (Fig. \ref{fig:emlnmap}), as well as the complex, non-axisymmetric brightness structure of the two hotspot regions (Fig. \ref{fig:emlnvlamap}). 
\par Since we infer mildly relativistic flow at the base of the radio jet, we argue that two distinct acceleration mechanisms exist in our radio source. 
\newline 

\text{1.} The first mechanism is governed by flow speeds, $\beta_{\rm j}$ $\gtrsim$ 0.5, and seems the likely dominant acceleration process in the jet base -- inferred from our lower limit of 0.5 for $\beta_{\rm j}$. Since knots and hotspots are analogous structures, it is plausible that strong relativistic shocks occur in both, particularly in the high-emissivity regimes. These shocks have the tendency to accelerate electrons to high Lorentz factors, allowing X-ray synchrotron emission in the hotspot regions comparable to that observed by \citet{Hardcastle05+} in the bright knots at the jet base. 
\newline 

\text{2.} The second mechanism dominates when the flow speed, $\beta_{\rm j}$ falls below about 0.5. In this regime, there is deceleration by entrainment such that the flow is transonic and dominated by weaker shocks. As pointed out by \citet{Rieger&Duffy04}, steady shear acceleration occurs under this condition. In addition to the jet base, we favour this as the likely dominant process at play in the extended structures (radio plumes) of 3C\,465, particularly as $\beta_{\rm j}$ approaches a limiting value $\ll$ 0.5 -- implying non-relativistic flows/shocks. We note that by this conclusion, we are by no means neglecting the likelihood that transverse velocity gradients must exist in these regions.

\par We infer particle acceleration in the plumes based on our radiative lifetime model and the observed continued collimated outflow of plasma from the site of particle injection (hotspots) into the plume materials. This compare with predictions from twin-beam model for radio sources (e.g. \citealt{Blanford&Rees74}), and as noted by \citet{Blanford&Eichler87}, these high energy electron populations, which are responsible for emission at such large distances must have been accelerated \textit{in situ}, most likely via weaker/non-relativistic shock acceleration as seen in the present study. We note here that it is still not clear whether or not there is any \textit{real} acceleration in this {\textquotedblleft{flow regime}\textquotedblright} or whether the spectral structures we see are just the result of newly accelerated particles mixing with the old ones in the plume material. 

\section{Conclusion and future work}

We have presented the highest resolution and sensitivity maps to date of the WAT source 3C\,465 jet, derived by combining radio data from e-MERLIN and VLA observations. Total intensity maps and a spectral index map (derived from two frequencies at 1.5-arcsec resolution) of the source are also presented. A comprehensive description of these maps and detailed spectral analysis to study the plausible underlying mechanisms of where and how particles are accelerated in the knots, hotspots and plumes of the radio source have also been presented. 
\par We obtain a sidedness ratio of 14.85 $\pm$ 0.80 and consequently derive lower and upper limit values of 0.5$c$ and 61$^{\circ}$ for the jet speed, $\beta_{\rm j}$ and angle to the line of sight, $\theta$ respectively. The principal results from our detailed study of the spectral index distribution are as follows:
\newline
 
\text{(i)} The spectral profile is fairly constant over almost the entire jet length with a mean value of <$\alpha_{\rm jet}$> = $-$0.71 $\pm$ 0.04, as normal from radio synchrotron emission; and that the spectral flattening within the first 4.4 kpc from the core coincides with the region hosting the bright knots, and is consistent with the sites of X-ray emission, and consequently high-energy electron acceleration at the base of the radio jet as observed by \citet{Hardcastle05+}. 
\par \text{(ii)} The spectral indices in the hotspot regions show little difference with mean values of <$\alpha_{\rm NWh}$> = $-$0.69 $\pm$ 0.01 and <$\alpha_{\rm SEh}$> = $-$0.65 $\pm$ 0.01, indicating that electron populations of similar properties are injected at these sites. Our spectral profiles suggest that the plumes are approximately homologous structures, in the sense that there is a clear trend in their spectra with downstream distance despite local variations in physical size and shape. We associate this difference in morphology to variations in the effects of the ICM in the two plumes. This may also account for the steeper spectrum in the NW plume, <$\alpha_{\rm NWp}$> = $-$1.43 $\pm$ 0.01 compared with the SE plume, <$\alpha_{\rm SEp}$> = $-$1.38 $\pm$ 0.01. There is also a clear tendency for the spectra to flatten slightly within the central regions of the plumes -- indicating the injection of a young electron population into the base of the plume materials, and plausible acceleration of particles at these sites. This is in agreement with our synchrotron lifetime model which suggests plausible reacceleration of particles within the plume materials.
\par \text{(iii)} First-order Fermi process at mildly relativistic shocks is the most probable acceleration mechanism at play in the radio source 3C\,465. Following earlier work by \citet{Laing&Bridle13}, our data are consistent with two acceleration mechanisms; (a) when bulk flow speeds, $\beta_{\rm j}$ $>$ 0.5, and (b) when flow speeds, $\beta_{\rm j}$ $<$ 0.5. The first case can accelerate electrons to high Lorentz factors, whereas the second scenario must occur at slower speeds and larger distances. 
\par For future work, we aim to observe over a broad range of frequencies to study in detail the \textit{deviations from power law spectra} which are indicators of synchrotron ageing and plausible diagnostics of the acceleration mechanism and as well, investigate the orientation and degree of ordering of magnetic field in the jet base.

\section*{Acknowledgements}

This research was supported by a Newton Fund project, DARA (Development in Africa with Radio Astronomy), and awarded by the UK's Science and Technology Facilities Council (STFC) - grant reference ST/R001103/1. We thank the anonymous referee for their prompt review and helpful comments. EBM would like to thank Javier Moldon (e-MERLIN Project Scientist) for assisting in the initial data reduction process of the e-MERLIN data at Jodrell Bank Centre for Astrophysics, University of Manchester. The National Radio Astronomy Observatory is a facility of the National Science Foundation operated under cooperative agreement by Associated Universities, Inc. The e-MERLIN is UK's national facility for radio astronomy and is operated by the University of Manchester from the Jodrell Bank Observatory (JBO) on behalf of STFC. This work has made use of the University of Hertfordshire's high-performance computing facility (\url{https://uhhpc.herts.ac.uk/}).

%%%%%%%%%%%%%%%%%%%%%%%%%%%%%%%%%%%%%%%%%%%%%%%%%%

%%%%%%%%%%%%%%%%%%%%%% REFERENCES %%%%%%%%%%%%%%%%%%%%

% The best way to enter references is to use BibTeX:

\bibliographystyle{mnras}
\bibliography{mnras_template.bib} % if your bibtex file is called example.bib

% Alternatively you could enter them by hand, like this:
% This method is tedious and prone to error if you have lots of references
%\begin{thebibliography}{99}

%\end{thebibliography}

%%%%%%%%%%%%%%%%%%%%%%%%%%%%%%%%%%%%%%%%%%%%%%%%%%

%%%%%%%%%%%%%%%%% APPENDICES %%%%%%%%%%%%%%%%%%%%%

%\appendix

%\section{Some extra material} If you want to present additional material which would interrupt the flow of the main paper, it can be placed in an Appendix which appears after the list of references.

%%%%%%%%%%%%%%%%%%%%%%%%%%%%%%%%%%%%%%%%%%%%%%%%%%

% Don't change these lines
\bsp	% typesetting comment
\label{lastpage}
\end{document}